\documentclass[aps,prc,superscriptaddress,twoside,twocolumn,nofootinbib,showpacs]{revtex4-1}
\usepackage{amsmath,amssymb}
\usepackage{mathtools}
\usepackage{bbold}
\usepackage[usenames, dvipsnames]{xcolor}
\usepackage{braket}
\usepackage{graphicx,bm}
\usepackage{slashed}
\usepackage{ulem}
\usepackage{etoolbox,siunitx}

\robustify\bfseries

\newcolumntype{d}{D{$.$}{$.$}{3.2}}

\newcommand*{\abs}[1]{\left|#1\right|}

\newcommand*{\NLK}{{N\Lambda K}}
\newcommand*{\XLK}{{\Xi\Lambda K}}

\allowdisplaybreaks

\begin{document}

\title{\boldmath $\bar{K} + N \to K + \Xi$ reaction and $S=-1$ hyperon resonances}

\author{Benjamin C. Jackson}%
\affiliation{Department of Physics and Astronomy, The University of Georgia,
Athens, GA 30602, USA}

\author{Yongseok Oh}%
\email{yohphy@knu.ac.kr}
\affiliation{Department of Physics, Kyungpook National University,
Daegu 702-701, Korea}
\affiliation{Asia Pacific Center for Theoretical Physics, Pohang,
Gyeongbuk 790-784, Korea}

\author{H. Haberzettl}%
\email{helmut@gwu.edu}
\affiliation{Institute for Nuclear Studies and Department of Physics,
The George Washington University, Washington, DC 20052, USA}

\author{K. Nakayama}%
\email{nakayama@uga.edu}
\affiliation{Department of Physics and Astronomy, The University of Georgia,
Athens, GA 30602, USA}
\affiliation{Institut f\"ur Kernphysik and Center for Hadron Physics,
Forschungszentrum J\"ulich, 52425 J\"ulich, Germany}

\date{\today}

\begin{abstract}
The $\bar{K} + N \to K + \Xi$ reaction is studied for center-of-momentum energies ranging 
from threshold to 3~GeV in an effective Lagrangian approach that includes the hyperon 
$s$- and $u$-channel contributions as well as a phenomenological contact amplitude. 
The latter accounts for the rescattering term in the scattering equation and possible short-range 
dynamics not included explicitly in the model. 
Existing data are well reproduced and three above-the-threshold resonances were found to be 
required to describe the data, namely, the $\Lambda(1890)$, $\Sigma(2030)$, and $\Sigma(2250)$. 
For the latter resonance we have assumed the spin-parity of $J^P=5/2^-$ and a mass of 2265~MeV. 
The $\Sigma(2030)$ resonance is crucial in achieving a good reproduction of not only the measured 
total and differential cross sections, but also the recoil polarization asymmetry. 
More precise data are required before a more definitive statement can be made about the other two
resonances, in particular, about the $\Sigma(2250)$ resonance that is introduced to describe a small 
bump structure observed in the total cross section of $K^- + p \to K^+ + \Xi^-$. 
The present analysis also reveals a peculiar behavior of the total cross section data in the threshold 
energy region in $K^- + p \to K^+ + \Xi^-$, where the $P$- and $D$-waves dominate instead of the 
usual $S$-wave. 
Predictions for the target-recoil asymmetries of the $\bar{K} + N \to K + \Xi$ reaction are also 
presented.
\end{abstract}


\maketitle

%%%%%%%%%%%%%%%%%%%%%%%%%%%%%%%%%%%%%%%%%%%

%%%%%%%%%%%%%%%%%%%%%%%%%%%%%%%%%%%%%%%%%%%

\section{Introduction}

Hadron spectroscopy is an essential part of the investigation to understand the
non-perturbative regime of Quantum Chromodynamics (QCD). In principle, an
ab-initio approach to hadron resonance physics can be provided by lattice QCD
simulations. In particular, the spectra of excited baryons observed in the
recent lattice simulations~\cite{HSC12,BGR13} hold the promise of explaining
the rich dynamics in the resonance energy region in the near future. Once quark
masses drop towards more reasonable values and finite volume effects are fully
under control, a close comparison to experimental data will be possible. Other
approaches such as the dynamical Dyson-Schwinger~\cite{WCCR12}, constituent
quark models~\cite{CR00,RM11}, and the Skyrme model~\cite{Oh07} also generate
resonance spectra. Unitarized Chiral Perturbation Theory also provides a
complementary picture of some of the low-lying resonances~\cite{OR09,MBM12}. To
compare these theoretical results with the experimental data, a reliable
reaction theory capable of identifying resonances and extracting the
corresponding resonance parameters is required. Such reaction theories, based
on a coupled-channel approach, have been
developed at various degrees of sophistication and are being improved~%
\cite{ABSW08,ABKN11,CKYDT07,TKCC10,SLM12,MSL06,KNLS10,RDHH12}.
So far, most of the experimentally extracted baryon resonances come from the pion-induced
reaction experiments, especially the $\pi N$ scattering, and about 16 nucleon resonances and
11 $\Delta$ resonances have been identified~\cite{PDG14}.
A number of $\Lambda$ and $\Sigma$ baryons, which are particles with strangeness quantum
number $S=-1$, have been also discovered~\cite{PDG14}.
A review on the status of baryon spectroscopy is given, e.g., in Ref.~\cite{KR09}.

Although the multi-strangeness baryons ($S < -1$) have played an important role
in the development of our understanding of strong interactions, and thus,
should be an integral part of any baryon spectroscopy program, the current
knowledge of these baryons is still extremely limited. In fact, the SU(3)
flavor symmetry allows as many $S=-2$ baryon resonances, called $\Xi$, as there
are $N$ and $\Delta$ resonances combined ($\sim 27$); however, until now, only
eleven $\Xi$ baryons have been discovered~\cite{PDG14}. Among them, only three
[ground state $\Xi(1318)1/2^+$, $\Xi(1538)3/2^+$, and $\Xi(1820)3/2^-$]
have their quantum numbers assigned.%
\footnote{The parity of the ground state $\Xi$ has not been measured explicitly
yet, but its assignment is based on quark models and SU(3) flavor symmetry. }
This situation is mainly due to the fact that multi-strangeness particle
productions have relatively low yields. For example, if there are no strange
particles in the initial state, $\Xi$ is produced only indirectly and the yield
is only of the order of nb in the photoproduction reaction~\cite{CLAS07b},
whereas the yield is of the order of $\mu$b~\cite{FMMR83b} in the hadronic
$\bar{K}$-induced reaction, where the $\Xi$ is produced directly because of the
presence of an $S=-1$ $\bar{K}$ meson in the initial state. The production
rates for $\Omega$ baryons with $S=-3$ are much lower~\cite{RHHK13}.

The study of multi-strangeness baryons has started to attract a renewed
interest recently. Indeed, the CLAS Collaboration at Thomas Jefferson National
Accelerator Facility (JLab) plans to initiate a $\Xi$ spectroscopy program
using the upgraded 12-GeV machine, and measure exclusive $\Omega$
photoproduction for the first time~\cite{VSC12}. Some data for the production
of the $\Xi$ ground state, obtained from the 6-GeV machine, are already
available~\cite{CLAS07b}. They were analyzed by some of the present
authors~\cite{NOH06,MON11} within an effective Lagrangian approach. J-PARC is
going to study the $\Xi$ baryons via the $\bar{K} + N \to K + \Xi$ process (which
is the reaction of choice for producing $\Xi$) in connection to its program
proposal for obtaining information on $\Xi$ hypernuclei spectroscopy. It also
plans to study the $\pi + N \to K + K + \Xi$ reaction as well as $\Omega$
production~\cite{Ahn06,Takahashi13}. At the FAIR facility of GSI, the reaction
$\bar{p} + p \to \bar{\Xi} + \Xi$ will be studied by the PANDA
Collaboration~\cite{PANDA09}. Quite recently, lattice QCD calculations of the
baryon spectra, including those of $\Xi$ and $\Omega$ baryons, have also been
reported, for example, in Refs.~\cite{HSC12,BGR13}.

In the present work, we concentrate on the production of $S=-2$ $\Xi$ and, in
particular, on the production reaction process of the ground state $\Xi$,
\begin{equation}
\bar K(q)+N(p)\rightarrow K(q')+\Xi(p') ~,
\label{reac}
\end{equation}
where the arguments indicate the corresponding particle's on-shell
four-momentum.
This reaction has been studied experimentally mainly throughout the 60's~%
\cite{PPSS62,CPSS64,BEHM66,BGLOR66,LRSY66,TS67,TFHL68,MB68,BMPT68,DBHM69},
which was followed by several measurements made in the 70's and 80's~%
\cite{SABRE71,DBRV72,CDJSE73,RBGP73,GGBP75,BGKS77,DKPO83}. The existing data
are rather limited and suffer from large uncertainties. The total cross section
and some of the differential cross section data are tabulated in
Ref.~\cite{FMMR83b}. We shall return to the discussion of these experimental
data later on. Early theoretical attempts to understand the above reaction are
very few and can be found in Refs.~\cite{EJ67,James67,ASNS71,MS82,DG83b}.
Recent calculations are reported by Sharov \textit{et al.\/}~\cite{SKL11} and
by Shyam \textit{et al.\/}~\cite{SST11}. The former authors have considered
both the total and differential cross sections as well as the recoil
polarization data in their analysis, while the latter authors have considered
only the total cross section data, although they too have predicted the
differential cross sections, mentioning that they found it difficult to use the
differential cross section data~\cite{DBHM69} for several reasons. Although the
analyses of Refs.~\cite{SKL11,SST11} are both based on very similar effective
Lagrangian approaches, the number of $S=-1$ hyperon resonances included in the
intermediate state are different. While in Ref.~\cite{SKL11} only the
$\Sigma(1385)$ and $\Lambda(1520)$ are considered in addition to the
above-threshold $\Sigma(2030)$ and $\Sigma(2250)$ resonances,\footnote{The
   production threshold energy for the reaction of Eq.~(\ref{reac}) is about
   $1813$ MeV.}
in Ref.~\cite{SST11} eight of the 3- and 4-star $\Lambda$ and $\Sigma$
resonances with masses up to 2.0 GeV have been considered. While the authors of
Ref.~\cite{SKL11} pointed out the significance of the above-threshold
resonances, the authors of Ref.~\cite{SST11} have found the dominance of the
sub-threshold $\Lambda(1520)$ resonance.
Reaction~(\ref{reac}) has been also considered quite recently by Magas
\textit{et al.}~\cite{MFR14} within the coupled channels Unitarized Chiral
Perturbation approach in connection to the issue of determining the parameters
of the next-to-leading-order interactions. The authors of Ref.~\cite{MFR14}
have added the $\Sigma(2030)$ and $\Sigma(2250)$ resonances into their
calculation to improve the fit quality to the total cross section data. Just
recently, the Argonne-Osaka group~\cite{KNLS14} reported applying their
Dynamical Coupled Channels approach to $\bar{K}$-induced two-body reactions for
center-of-momentum (c.m.) energies up to $W = 2.1$~GeV. In the reported work,
both the total and differential cross sections were calculated, but the
extracted resonance parameter values are not yet available.

We note here that the proper identification of resonances and the reliable
extraction of their parameters requires detailed knowledge of the analytic
structures of the scattering amplitude that, to date, can only be obtained
through a full coupled-channel treatment, such as that of Ref.~\cite{KNLS14}.
However, because the currently available data in the $K\Xi$ channel are scarce
and of low quality, they do not provide sufficient constraints for the model
parameters to permit an in-depth analysis of that channel~\cite{KNLS14}. In
this context, we mention that a coupled-channel partial-wave analysis of
$\bar{K}$-induced reactions up to $W=2.1$~GeV has also been performed recently
by the Kent State University group~\cite{ZTSM13a,ZTSM13b} which includes the
$\bar{K}N$, $\pi\Lambda$, $\pi\Sigma$, $\pi\Lambda(1520)$, $\pi\Sigma(1385)$,
$\bar{K}^*N$, and $\bar{K}\Delta$ channels, but not the $K\Xi$ channel.

Some of the model-independent aspects of the reaction~(\ref{reac}) have been
studied recently by the present authors~\cite{NOH12,JOHN14}. In the present
work, we perform a model-dependent analysis of the existing data between
threshold and a c.m. energy of 3~GeV based on an effective Lagrangian approach
that includes a phenomenological contact amplitude which accounts for the
rescattering contributions and/or unknown (short-range) dynamics that have not
been included explicitly into the model. While the tree-level model presented
here is not very sophisticated, it captures the essential aspects of the
process in question. As such, the use of a simplified, yet efficient model is
particularly well suited for a situation, such as for the
reaction~(\ref{reac}), where scarce and poor data prevent a more detailed and
complete treatment. The present study is our first step toward building a more
complete reaction model capable of reliably extracting the properties of
hyperons from the forthcoming experimental data, in addition to providing some
guidance for planning future experiments.
One of the purposes of the present work is to search for a clearer
evidence of the $S=-1$ hyperon resonances in reaction (\ref{reac}).
However, we emphasize that our main interest here lies not so much in the
accurate extraction of $S=-1$ hyperon resonance parameters, but in
an exploratory study to learn about the pertinent reaction
mechanisms and, in particular, to identify the resonances that come out to be
most relevant for the description of the existing $\Xi$ production data. In
fact, with the exception of the $\Sigma(2250)$ resonance, whose mass was
adjusted slightly to better reproduce the observed bump structure in the total
cross section in the charged $\Xi$ production, the masses and widths of the
resonances incorporated here are taken from other sources, as explained in
Sec.~\ref{sec:results} below. Only the product of the coupling constants and
the cutoff parameters in the corresponding form factors are adjusted in the
present work.

The investigation of  reaction (\ref{reac}) also impacts the study of $\Xi$
hypernuclei, where the elementary process of Eq.~(\ref{reac}) is an input for
the models of hypernuclei productions~\cite{DG83b,YMFTI94,TKA95,KH10}. As
mentioned before, there is a proposed program at J-PARC and eventually at
GSI-FAIR to obtain information about the spectroscopy of $\Xi$ hypernuclei
through the antikaon-induced reactions on nuclear targets. Establishing the
existence and properties of $\Xi$ hypernuclei is of considerable importance for
a number of reasons and the study of  reaction (\ref{reac}) is an essential
step to this end.

The present paper is organized as follows. In Sec.~\ref{sec:model}, our model
for describing reaction~(\ref{reac}) is presented, with some technical details
supplied in the Appendix. In Sec.~\ref{sec:results}, the results of our model
calculations are presented and discussed. Section~\ref{sec:conclusion} contains
our summary and conclusions.

%%%%%%%%%%%%%%%%%%%%%%%%%%%%%%%%%%%%%%%%%%%

\section{Model Description} \label{sec:model}

\begin{figure}[t]
\includegraphics[width=0.75\columnwidth,clip=]{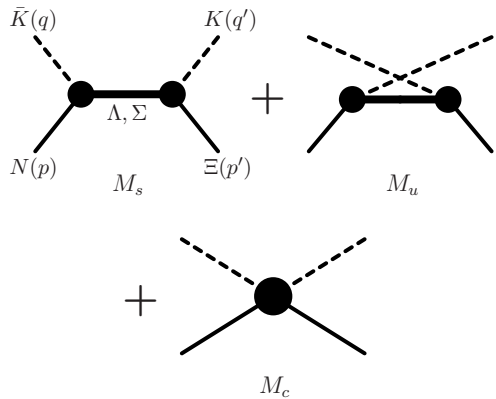}
\caption{\label{fig:diagrams}%
Diagrams describing the amplitude (\ref{eq:BS-approx}) in the present
calculation. The labeling of the external legs of the $s$-channel diagram,
$M_s$, follows the reaction equation  (\ref{reac}); the labels apply
correspondingly also to the external legs of the $u$-channel diagram, $M_u$,
and the contact term $M_c$. The intermediate hyperon exchanges, $\Lambda$ and
$\Sigma$, indicated for $M_s$ also appear in $M_u$. The details of the contact
amplitude, $M_c$, are discussed in Sec.~\ref{sec:model}.}
\end{figure}

The reaction amplitude, $T$, describing a two-body process like the reaction
(\ref{reac}) is, in general, given by the Bethe-Salpeter equation,
\begin{equation}
T = V + VG_0T \, ,
\label{eq:BS0}
\end{equation}
where $V$ stands for the (two-body) meson-baryon irreducible (Hermitian) driving amplitude
and $G_0$ describes free relative meson-baryon motion.
Note that the above equation represents, in principle, a coupled-channels equation in
meson-baryon channel space.
It can be recast into the pole and the non-pole parts as
\begin{equation}
T = T^{\rm P} + T^{\rm NP} \, ,
\label{eq:BS1}
\end{equation}
where the non-pole part $T^{\rm NP}$ obeys
\begin{equation}
T^{\rm NP} = V^{\rm NP} + V^{\rm NP} G_0 \, T^{\rm NP}
\label{eq:BS-NP}
\end{equation}
with
\begin{equation}
V^{\rm NP} \equiv V - V^{\rm P}
\label{eq:V-NP}
\end{equation}
denoting the one-baryon irreducible (non-pole) part of the driving amplitude, $V$.
Here, $V^{\rm P}$ stands for the one-baryon reducible (pole) part of $V$ in the form of
\footnote{%
  The bra and ket notation here is used only as a quick visual cue to identify
  incoming and outgoing vertices, respectively. They are not to be taken as
  Hilbert space states in the usual sense.}
\begin{equation}
V^{\rm P} = \sum_r \ket{F_{0 r}} S_{0 r} \bra{F_{0 r}}  ,
\label{eq:V-P}
\end{equation}
where $\ket{F_{0 r}}$ and $S_{0 r} = (p^2_r - m^2_{0 r} + i0)^{-1}$ stand for
the so-called bare vertex and bare baryon propagator, respectively. The
summation runs over the baryons in the intermediate state, each specified by
the index $r$. The four-momentum and the bare mass of the propagating baryon
are denoted by $p_r^{}$ and $m_{0 r}^{}$, respectively.
As can be seen in Fig.~\ref{fig:diagrams}, $V^{\rm P}$ is the
sum of the $s$-channel Feynman diagrams corresponding to bare
baryon propagations in the intermediate state. The pole part of the reaction amplitude $T^{\rm P}$
in Eq.~(\ref{eq:BS1}) is given by
\begin{equation}
T^{\rm P} = \sum_{r'r} \ket{F_{r'}} S_{r'r} \bra{F_r} ,
\label{eq:BS-P}
\end{equation}
where the so-called dressed vertex reads
\begin{align}
\ket{F_{r'}} & = \left( 1 + T^{\rm NP} G_0 \right) \ket{F_{0 r'}}  , \nonumber \\
\bra{F_r} & = \bra{F_{0 r}} \left( 1 + G_0 \, T^{\rm NP} \right)  ,
\label{eq:Dres-vertx}
\end{align}
and the dressed propagator $S_{r'r}$ is written as
\begin{equation}
S^{-1}_{r'r} = S^{-1}_{0 r} \delta_{r'r}^{}  - \Sigma_{r'r} \,  ,
\label{eq:Dres-prop}
\end{equation}
with
\begin{equation}
\Sigma_{r'r} = \bra{F_{0 r'}} G_0 \ket{F_r}
\label{eq:Self-eneg}
\end{equation}
denoting the self-energy.
%

%%%%%%%%%%%%%%%%%%%%%%%%%%%%%%%%%%%%%%%%%%%%%%
%    Figure 1
%%%%%%%%%%%%%%%%%%%%%%%%%%%%%%%%%%%%%%%%%%%%%%

In the present work we shall make the following approximations to the reaction
amplitude in Eq.~(\ref{eq:BS1}). First, we approximate the pole part of the
reaction amplitude $T^{\rm P}$ by the $s$-channel Feynman amplitude, $M_s$,
specified by effective Lagrangians and phenomenological Feynman propagators.
Here, the dressed resonance coupling constants, dressed masses as well as the
corresponding widths are parameters either fixed from independent sources or
adjusted to reproduce the experimental data. The meson-baryon-baryon vertices
are obtained from the effective Lagrangians given in the Appendix; the
phenomenological Feynman propagators for dressed baryons are also found there.
Note that, here, the resonance couplings in the dressed propagators are
ignored.

Second, the non-pole part of the reaction amplitude $T^{\rm NP}$ is approximated as follows.
\begin{enumerate}
\item[(i)] Since there is no meson-exchange $t$-channel process in the
    present reaction, unless the exchanged meson is an exotic one with
    strangeness quantum number $S=2$, $V^{\rm NP}$  of the reaction is
    approximated by the $u$-channel Feynman amplitude, $M_u$, constructed
    from the same effective Lagrangians and Feynman propagators used to
    construct the $s$-channel Feynman amplitudes.

\item[(ii)] The rescattering term $V^{\rm NP}G_0\, T^{\rm NP}$ in $T^{\rm
    NP}$ of Eq.~(\ref{eq:BS-NP}) and other effects not explicitly included in
    the present approach are accounted for by a phenomenological contact
    term, $M_c$, which is specified below. This contact term will be
    discussed in more detail later.
\end{enumerate}

With the approximations described above, the reaction amplitude in the present work is
given by
\begin{equation}
T = M_s + M_u + M_c \,  ,
\label{eq:BS-approx}
\end{equation}
where $M_s$ and $M_u$ are the amplitudes from the $s$- and $u$-channel Feynman
diagrams, respectively; both amplitudes include the ground-state hyperons as
well as some of the $S=-1$ hyperon resonances in the intermediate state.
Figure~\ref{fig:diagrams} shows a diagrammatic representation of $M_s$, $M_u$,
and $M_c$.

The amplitude for the reaction of Eq.~(\ref{reac}) can be decomposed into
spin-non-flip and spin-flip contributions.
Their respective partial-wave decomposed forms read%
   \footnote{There are in total four spin matrix elements to describe the reaction (\ref{reac}).
   However, only two of them, corresponding to the spin-non-flip and spin-flip processes, are
   independent due to the reflection symmetry about the reaction plane for parity conserving
   processes. See Ref.~\cite{JOHN14} for more detailed discussions. }
\begin{subequations}\label{eq:PW}
\begin{align}
M_{++} & = M_{--}  \nonumber \\
& =  \frac{1}{4\pi} \sum_{L,T} \left[ (L+1) \, M^{TJ_+}_{L}(p', p) + L\, M^{TJ_-}_{L}(p', p)\right]
\nonumber \\ & \qquad\qquad \times
P_{L}^{}(\hat{\bm{p}} \cdot \hat{\bm{p}}') \, \hat{P}_T^{}  \, ,
 \\[1ex]
M_{+-} & = - M_{-+}  \nonumber \\
& =  \frac{1}{4\pi} \sum_{L,T} \left[ M^{TJ_+}_{L}(p', p) - M^{TJ_-}_{L}(p', p) \right]
\nonumber \\ & \qquad\qquad \times
P^1_{L}(\hat{\bm{p}} \cdot \hat{\bm{p}}') \, \hat{P}_T^{}  \, ,
\end{align}
\end{subequations}
where initial and final momenta are as in Fig.~\ref{fig:diagrams}. The indices
$s', s = \pm$ in $M_{s's}$ stand for spin-up ($+$) and spin-down ($-$) of the
final ($s'$) and initial ($s$) states quantized along the incoming
momentum direction $\hat{\bm{p}}$, and $J_\pm \equiv L\pm\frac12$ (for $L=0$,
the corresponding $J_-$ terms are zero). $M^{TJ_\pm}_L$ are diagonal elements
of the more general partial-wave amplitudes introduced in
Ref.~\cite{JOHN14} (where full technical details can be found).
The Legendre and associated Legendre functions are denoted by $P_{L}(x)$ and
$P^1_{L}(x)$, respectively,%
   \footnote{Here, the phase convention for the associated Legendre function is
   such that $P^1_{1}(x) = \sin(x)$.}
with argument $\hat{\bm{p}} \cdot \hat{\bm{p}}'=\cos\theta$, where $\theta$ is
the scattering angle. The total angular momentum, orbital angular momentum, and
total isospin of the meson-baryon state are represented by $J, L$, and
$T$, respectively. $\hat{P}_T^{}$ stands for the isospin projection operator
onto the total isospin 0 or 1 as $T=0$ or $T=1$, respectively. Explicitly,
$\hat{P}_{T=0}^{} = (3 + \bm{\tau}_1^{} \cdot \bm{\tau}^{}_2) / 4$ and
$\hat{P}_{T=1}^{} = (1 - \bm{\tau}_1^{} \cdot \bm{\tau}_2^{}) / 4$.

The phenomenological contact amplitude, $M_c$, is now decomposed in terms of
spin amplitudes similar to Eqs.~(\ref{eq:PW}) as well. Following the essential
idea of Ref.~\cite{NL05}, the corresponding contact term contributions are
parameterized as
\begin{subequations}\label{ampl_cont}
\begin{align}
M_{c\,++} & = M_{c\,--}  \nonumber \\
& = \sum_{L,T} g_1^{LT} \left( \frac{p'}{\Lambda_S} \right)^{L}
\exp\left(-\alpha^{T}_{L} \frac{p'^2}{\Lambda^2_S}\right)
% \nonumber\\
%  &\qquad\qquad\times
P_{L}^{}(\hat{\bm{p}} \cdot \hat{\bm{p}}')
\hat{P}_T^{} \, ,
  \\[1ex]
M_{c\,+-} & =- M_{c\, -+} \nonumber \\
& = \sum_{L,T} g_2^{LT} \left( \frac{p'}{\Lambda_S} \right)^{L}
\exp\left(-\beta^{T}_{L} \frac{p'^2}{\Lambda^2_S}\right)
% \nonumber\\
%  &\qquad\qquad\times
 P^1_{L}(\hat{\bm{p}} \cdot \hat{\bm{p}}')
 \hat{P}_T^{} \, ,
\end{align}
\end{subequations}
with $g_1^{LT}\equiv a^T_L\exp\left(i\phi_a^{TL}\right)$, $g_2^{LT} \equiv
b^{T}_{L}\exp\left(i\phi_b^{TL}\right)$, and $\alpha^{T}_{L}$, $\beta^{T}_{L}$
being constants to be fitted. $\phi_x^{TL}$ for $x=a,b$ is the complex phase
angle parameter which renders the contract amplitude, $M_c$,  complex and
$\Lambda_S$ is a typical scale parameter of the reaction at hand. The momentum
dependence of the partial-wave matrix elements given above is particularly well
suited for hard processes, which have a large momentum transfer and whose
amplitudes are expected to be independent of energy and nearly constant apart
from the centrifugal barrier effects.
Though reaction (\ref{reac}) is not a very hard process,%
     \footnote{For example, the momentum transfer of this reaction at threshold
      is about 200 MeV.}
the $p'^{L}$-dependence nonetheless captures the essence of the behavior of the
amplitude at low momentum in the final state. For further details, we refer to
Ref.~\cite{NL05}. The exponential factor in Eq.~(\ref{ampl_cont}) is simply a
damping factor to suppress the high momentum behavior introduced by $p'^{L}$.

It should be noted that our phenomenological contact term, $M_c$, can only
account for effects with a smooth energy dependence. Effects from, for example,
dynamically generated resonances and/or channel
couplings~\cite{DN09,RO13,RTHK12,RHKT14}, etc., that exhibit a strong variation
of the amplitude as a function of energy cannot be described by the contact
term.

The amplitudes  $M_s$, $M_u$, and $M_c$ must be added up to obtain the total
scattering amplitude of Eq.~(\ref{eq:BS-approx}). To ensure that there is no
ambiguity in the relative phase of $M_s+M_u$ and $M_c$ caused by different
Feynman rules, we give an explicit calculation of $M_s$ and $M_u$ for
$\Lambda(1116)$ in the Appendix.

%%%%%%%%%%%%%%%%%%%%%%%%%%%%%%%%%%%%%%%%%%%%%%
%    Table (I)
\begin{table*}[t]
\centering
\caption{The $\Lambda$ and $\Sigma$ hyperons listed by the Particle Data Group~\cite{PDG14}
(PDG) as three- or four-star states.
The decay widths and branching ratios of higher-mass resonances ($m_r^{} > 1.6$ GeV) are in a
broad range, and the coupling constants are determined from their centroid values.
In the present work, the masses ($m_r$) and widths ($\Gamma_r$) of the hyperons as given in this
table have been used, except for the $\Sigma(2250)$ resonance.
For the latter resonance, see the text.}
\begin{tabular}{c@{\extracolsep{1em}}S[table-format=1.1]S[table-format=3.0]cS[table-format= 1.1]|cS[table-format=4.0]S[table-format= >1.1]cS} \hline\hline
\multicolumn{5}{c|}{$\Lambda$ states} & \multicolumn{5}{c}{$\Sigma$ states} \\
\hline State & {$m_r$ (MeV)} & {$\Gamma_r$ (MeV)} & Rating & {$|g_{N\Lambda K}^{}|$}
& State & {$m_r$ (MeV)} & {$\Gamma_r$ (MeV)} & Rating  & {$|g_{N\Sigma K}^{}|$} \\ \hline
$\Lambda(1116)$ $1/2^+$ & 1115.7   &  & **** &     &
  $\Sigma(1193)$ $1/2^+$ & 1193 &              & **** &     \\
$\Lambda(1405)$ $1/2^-$ & 1406  & 50 & **** &       &
  $\Sigma(1385)$ $3/2^+$ & 1385  & 37 & **** &      \\
$\Lambda(1520)$ $3/2^-$ & 1520  & 16 & **** &       &
                 &         &                   &       \\
\hline
$\Lambda(1600)$ $1/2^+$ &1600   & 150 & *** &  4.2 &
  $\Sigma(1660)$ $1/2^+$ & 1660  & 100 & *** & 2.5  \\
$\Lambda(1670)$ $1/2^-$ & 1670  & 35 & **** &  0.3 &
  $\Sigma(1670)$ $3/2^-$ & 1670  & 60 & **** & 2.8 \\
$\Lambda(1690)$ $3/2^-$ & 1690 & 60 & **** &  4.0 &
  $\Sigma(1750)$ $1/2^-$ & 1750  & 90 & *** & 0.5 \\
$\Lambda(1800)$ $1/2^-$ & 1800  & 300 & *** &  1.0 &
  $\Sigma(1775)$ $5/2^-$ & 1775  & 120 & **** & \\
$\Lambda(1810)$ $1/2^+$ & 1810  & 150 & *** &  2.8 &
  $\Sigma(1915)$ $5/2^+$ & 1915  & 120 & **** & \\
$\Lambda(1820)$ $5/2^+$ & 1820  & 80 & **** &      &
  $\Sigma(1940)$ $3/2^-$ & 1940  & 220 & *** & < 2.8  \\
$\Lambda(1830)$ $5/2^-$ & 1830  & 95 & **** &      &
  $\Sigma(2030)$ $7/2^+$ & 2030  & 180 & **** & \\
$\Lambda(1890)$ $3/2^+$ & 1890  & 100 & **** & 0.8 &
  $\Sigma(2250)$ $\ \ \ \ ?^?$
   & 2250 & 100 & ***  &  \\
$\Lambda(2100)$ $7/2^-$ & 2100  & 200 & **** & &  & & & \\
$\Lambda(2110)$ $5/2^+$ & 2110  & 200 & *** & &  &  & & \\
$\Lambda(2350)$ $9/2^+$ & 2350  & 150 & *** &  & & & & \\
\hline\hline
\end{tabular}
\label{tbl:hyperons}
\end{table*}
%%%%%%%%%%%%%%%%%%%%%%%%%%%%%%%%%%%%%%%%%%%%%%

Standard effective Lagrangian approaches include tree-level $s$-, $u$- and
$t$-channel diagrams, without phenomenological contact terms. Apart from
crossing symmetry demands, the inclusion of the $u$-channel amplitude, $M_u$,
in particular, is necessary to reproduce the backward peaking of the
differential cross sections. (See Sec.~\ref{sec:results}.) In fact, there are a
number of $\Lambda$ and $\Sigma$ resonances (cf.\ Table~\ref{tbl:hyperons})
that may contribute to this reaction. However, it happens that the $u$-channel
resonance contributions, especially from many of the sub-threshold resonances,
also give rise to a total cross section which keeps increasing with energy in
the present reaction process. This feature is not supported by the data, which
reach a peak and then fall off as a function of energy. Thus, one needs a
dynamical mechanism to suppress this rise in energy.

The effective Lagrangian approaches of Refs.~\cite{SKL11,SST11} have introduced
phenomenological mechanisms for dealing with this problem that are very similar
in spirit albeit somewhat different in technical detail. In both approaches,
the rise of the $u$-channel resonance diagrams was suppressed with functions
that smoothly cut off their contributions at high
energies.%
    \footnote{%
    According to a private communication by one of the authors of Ref.~\cite{SST11}, the form factor
    given in Eq.~(3) of that work only applies to the $s$-channel; the $u$-channel was suppressed
    instead by the form factor given in Eq.~(5) of Ref.~\cite{SS08}.}
While the respective procedures generally provide satisfactory agreement with the data,
they both violate crossing symmetry even at the tree-level.

In our model calculations, we also see the same undesirable rise of $u$-channel
contributions if we leave out contact terms. We interpret this to mean that the
rescattering term $V^{\rm NP} G_0\,T^{\rm NP}$ of the non-pole $T$-matrix in
Eq.~(\ref{eq:BS-NP}) would be responsible for providing the cancellation for
the increasing $u$-channel resonance amplitudes. We account here
phenomenologically for these in detail very complex dynamics by introducing
contact terms, and our results in Sec.~\ref{sec:results} will show that this
will indeed allow us to treat both $s$- and $u$-channel contributions
consistently, and at the same time avoid the high-energy $u$-channel contributions.

In general, it seems that the problem has two scales, corresponding to
long-range and short-range dynamics. The latter is, of course, sensitive to the
form factors used at the meson-baryon vertices to account for the composite
nature of the hadrons, and the use of phenomenological contact terms seems to
be warranted to account for additional structure effects. Problems with two
scales have been addressed in the past, where some authors have introduced two
form factors, one soft and other hard, to mimic such effects~\cite{PFG94}.
Also, in effective field theories the unknown short-range dynamics is accounted
for by contact terms.

%%%%%%%%%%%%%%%%%%%%%%%%%%%%%%%%%%%%%%%%%%%

\section{Results} \label{sec:results}

In this section, we present our results for the reaction $\bar K+N \to K+\Xi$
in different isospin channels. More specifically, we investigate the reactions
$K^- + p \to K^+ + \Xi^-$, $K^- + p \to K^0 + \Xi^0$, and $K^- + n \to K^0 +
\Xi^-$ considering all the available data on the total and differential cross
sections as well as recoil polarization asymmetries.

Before we present our results, we briefly remark on the experimental data
considered in this work, i.e., total cross sections, differential cross
sections, and recoil polarization asymmetries. These data come from different
sources~\cite{BGLOR66,BEHM66,LRSY66,TS67,TFHL68,DBHM69,CDJSE73,BMPT68} and are
available in various forms. Some of them are not in the tabular (numerical)
form that can be readily used but are given only in graphical form or as
parametrization in terms of the Legendre polynomial expansions. In
Ref.~\cite{SKL11}, Sharov \textit{et al.\/} have carefully considered the data
extraction from these papers. We have checked that the extracted data are
consistent with those in the original papers within the permitted accuracy of
the check. In the present work, we use these data, and no cross sections
resulting from the expansion coefficients are considered here.

As mentioned before, there are a number of 3- and 4-star $\Lambda$ and $\Sigma$
resonances, including those low-mass sub-threshold ones that contribute, in
principle, to reaction~(\ref{reac}). A list of these hyperon resonances and
some of their properties is shown in Table~\ref{tbl:hyperons}. However, apart
from the ground state $\Lambda(1116)$ and $\Sigma(1193)$, the required
information for most of these resonances on the resonance parameters, such as
the coupling strength (including their signs) to $\Xi$ and/or $N$, are largely
unknown. Therefore, the strategy adopted in this work is to consider these
parameters as fit parameters and consider the minimum number of resonances
required to reproduce the existing data. In particular, we have considered only
those resonances that give rise to a considerable contribution to the cross
section within a physically reasonable range of the resonance parameter values.
More specifically, during the fitting procedure, resonances were added one by
one to the model and the quality of fit was checked. It should be mentioned
that we have also checked the influence of various combinations of resonances
at a time (and not just one by one) to the fit quality. The resonances kept in
the presented calculation were those that increased the quality of the fit by a
noticeable amount with the variation in $\chi^2$ per data points $N$, namely,
$\delta\chi^2/N > 0.1$. An example of this procedure is shown in
Table~\ref{TABLE_res_chi2} where the results of adding one more resonance to
the current model, as specified later, is shown. We see that some of these
resonances improve the fit quality of the total cross section but not the other
observables or even worsen the fit quality slightly. We have not included these
resonances into our model because the total cross sections suffer from
relatively large uncertainties.

%%%%%%%%%%%%%%%%%%%%%%%%%%%%%%%%%%%%%%%%%%%%%%
%    Table (II)
\begin{table*}[t]
\caption{
Variation in $\chi^2$ per data point $N$, $\delta\chi^2/N$, obtained
when adding one more resonance to the current model (specified in
Table.~\ref{tab:para_g_contact}). A negative $\delta\chi^2/N$ corresponds to an
improvement in the result. The quantity $\delta\chi^2_i/N_i$ corresponds to
$\delta\chi^2/N$ evaluated for a given type of observable specified by index
$i = \sigma$(total cross section), $d\sigma$(differential cross section),
and $P$(recoil asymmetry). $N = N_\sigma + N_{d\sigma} + N_P$ denotes the
total number of data points. Furthermore, $\delta\chi^2_i/N_i$ is given for the
charged $\Xi^-$ ($\delta\chi^2_-/N_-$) and neutral $\Xi^0$
($\delta\chi^2_0/N_0$) production processes, separately. The last column
corresponds to $\delta\chi^2/N$ of the global fit considering all the data of
both reaction processes. The last row  corresponds to $\chi^2_i/N_i$ of the
current model. }

\begin{tabular}{c | S S S S | S S S S S | S}	\hline
  & \multicolumn{3}{c}{$\bar{K}^- + p \to K^+ + \Xi^-$} & & \multicolumn{4}{c}{$\bar{K}^- + p \to K^0 + \Xi^0$} &
  & \\ \hline
$Y$ added & {\ \ $\delta\chi_\sigma^2/N_\sigma$\ \ } &
{\ \ $\delta\chi_{d\sigma}^2/N_{d\sigma}$\ \ } & {\ \ $\delta\chi_P^2/N_P$\ \ } &
{\ \ $\delta\chi_-^2/N_-$\ \ } & & {\ \ $\delta\chi_\sigma^2/N_\sigma$\ \ } &
{\ \ $\delta\chi_{d\sigma}^2/N_{d\sigma}$\ \ } & {\ \ $\delta\chi_P^2/N_P$\ \ } &
{\ \ $\delta\chi_0^2/N_0$\ \ } & {\ \ $\delta\chi^2/N$\ \ } \\ \hline
%none	          	     & 1.54 & 1.73 & 2.25 & 1.76 & & 0.93 & 1.10 & 1.75 & 1.14 & 1.58 \\ \hline
$\Lambda(1405)$ & -0.01 & 0.03 & 0.00 & -0.01 & & 0.03 & 0.00 & 0.02 & 0.01 & 0.00  \\
$\Lambda(1600)$ & -0.02 & 0.00 & -0.01 & -0.01 & & 0.02 & 0.00 & 0.02 & 0.01 & 0.00  \\
$\Lambda(1670)$ & -0.01 & 0.00 & 0.00 & 0.00 & & 0.02 & 0.00 & 0.02 & 0.01 & 0.00  \\
$\Lambda(1800)$ & 0.00 & 0.01 & 0.00 & 0.00 & & -0.01 & 0.00 & 0.01 & 0.00 & 0.00  \\
$\Lambda(1810)$ & -0.01 & -0.01 & 0.00 & -0.01 & & 0.02 & 0.00 & 0.02 & 0.01 & 0.00 \\
$\Lambda(1520)$ & -0.06 & 0.02 & 0.00 & 0.00 & & -0.05 & -0.01 & 0.00 & -0.02 & 0.00  \\
$\Lambda(1690)$ & 0.00 & 0.00 & 0.00 & 0.00 & & 0.00 & 0.00 & 0.00 & 0.00 & 0.00  \\
$\Lambda(1820)$ & -0.08 & 0.01 & 0.01 & 0.00 & & -0.07 & 0.00 & -0.02 & -0.02 & -0.01  \\
$\Lambda(1830)$ & -0.05 & 0.01 & 0.01 & 0.00 & & 0.00 & 0.02 & 0.02 & 0.01 & 0.00  \\
$\Lambda(2110)$ & -0.02 & 0.02 & 0.01 & 0.01 & & -0.03 & -0.01 & -0.03 & -0.02 & 0.00  \\
$\Lambda(2100)$ & -0.08 & 0.04 & 0.03 & 0.02 & & -0.04 & -0.02 & -0.01 & -0.03 & 0.01  \\
$\Sigma(1660)$ & -0.02 & 0.00 & 0.00 & 0.00 & & -0.01 & 0.01 & 0.00 & 0.01 & 0.00  \\
$\Sigma(1750)$ & -0.01 & 0.01 & 0.00 & 0.00 & & -0.01 & 0.01 & 0.00 & 0.00 & 0.00  \\
$\Sigma(1670)$ & -0.01 & 0.00 & -0.01 & 0.00 & & 0.02 & 0.01 & 0.01 & 0.01 & 0.00  \\
$\Sigma(1940)$ & 0.02 & 0.00 & 0.01 & 0.00 & & 0.01 & -0.01 & 0.01 & -0.01 & 0.00  \\
$\Sigma(1775)$ & -0.01 & 0.01 & 0.04 & 0.01 & & -0.02 & 0.00 & -0.02 & -0.01 & 0.00  \\
$\Sigma(1915)$ & 0.01 & -0.01 & 0.00 & -0.01 & & -0.03 & 0.00 & 0.00 & -0.01 & -0.01  \\ \hline\hline
  & { $\chi_\sigma^2/N_\sigma$ } &
{ $\chi_{d\sigma}^2/N_{d\sigma}$ } & { $\chi_P^2/N_P$
} & { $\chi_-^2/N_-$ } & & { $\chi_\sigma^2/N_\sigma$ } &
{ $\chi_{d\sigma}^2/N_{d\sigma}$ } & { $\chi_P^2/N_P$} & { $\chi_0^2/N_0$ } &
{  $\chi^2/N$ } \\ \hline
       	     & 1.53 & 1.64 & 1.89 & 1.65 & & 0.88 & 1.06 & 1.74 & 1.10 & 1.49 \\ \hline
\end{tabular}
\label{TABLE_res_chi2}
\end{table*}
%
%%%%%%%%%%%%%%%%%%%%%%%%%%%%%%%%%%%%%%%%%%%%%%

Whenever appropriate, for each resonance considered in this work, the
corresponding coupling constants $g_{KYN}^{}$ and $g_{KY\Xi}^{}$ were
constrained in such way that the sum of the branching ratios $\beta_{Y \to KN}
+ \beta_{Y \to K\Xi}$ not to exceed unity. Because, within our model, the data
are sensitive only to the product of the coupling constants $g_{KYN}^{}
g_{KY\Xi}^{}$, setting $\abs{g_{KYN}^{}}=\abs{ g_{KY\Xi}^{}}$ for the purpose
of estimating the individual branching ratios, and only for this purpose, is a
simple way of keeping our coupling constant values within a physically
acceptable range. Admittedly, the currently existing data are limited and
suffer from large uncertainties, thus an accurate determination of the
resonance parameters are not possible at this stage. For this, one needs to
wait for new more precise data, possibly including more spin polarization data.
In this regard, the multi-strangeness baryon spectroscopy program using the
antikaon beam at J-PARC  will be of particular relevance. For the ground states
$\Lambda(1116)$ and $\Sigma(1193)$, the corresponding coupling constants are
estimated based on the flavor SU(3) symmetry relations~\cite{NOH06}.

It should be mentioned that, in principle, the coupling constants $g_{KYN}^{}$ and $g_{KY\Xi}^{}$
are complex quantities owing to the dressing mechanism of the resonance vertex as given by
Eq.~(\ref{eq:Dres-vertx}).
In the present work, we restrict them to be pure real to reduce the number of free parameters and
for the sake of simplicity. Note that the complex phases of the coupling constants are not arbitrary
in that they are constrained by unitarity of the scattering amplitude~\cite{DKT07}, a feature that is
absent in the amplitude based on a tree-level approximation as mentioned in the introduction.
However, we found that fit quality was not improved when we ignored theoretical constraints and
simply allowed for complex phases in the coupling constants.
A simple and proper way of accounting for unitarity within an effective Lagrangian approach, such as
the present one, is being developed and will soon be available elsewhere~\cite{RN15}.

The phenomenological contact amplitude $M_c$ contains two sets of free parameters,
$\{g_1^{LT},\alpha^T_{L}\}$ and $\{g_2^{LT},\beta^T_{L}\}$, to be fixed by adjusting to reproduce the
experimental data, for a given set of $\{L,T\}$ as shown in Eq.~(\ref{ampl_cont}).
In order to reduce the number of free parameters, we have assumed the parameter $\alpha^{T}_{L}$
to be equal to $\beta^T_{L}$ and independent on $T$ and $L$, i.e.,
$\alpha^{T}_{L}=\beta^T_{L}=\alpha$.
The scale parameter $\Lambda_S$ has been fixed as $\Lambda_S = 1~\mbox{GeV}$.
Note that the phenomenological contact amplitude can and should be complex in principle, since
it accounts for the rescattering contribution ($V^{\rm NP}G\, T^{\rm NP}$) of the non-pole $T$-matrix
which is complex in general.
Accordingly, the coupling strength parameters $g_1^{LT}$ and $g_2^{LT}$ are complex quantities.
In order to reduce the number of free parameters, we take their phases to be independent on $L$ and
$T$, so that, $\phi_a^{TL} = \phi_a$ and $\phi_b^{TL} = \phi_b$ for all sets $\{L,T\}$.
Also, in the present calculation, we find that it suffices to consider partial waves up to $L=2$ in the
contact amplitude to reproduce the existing data.

%%%%%%%%%%%%%%%%%%%%%%%%%%%%%%%%%%%%%%%%%%%%%%
%    Table (III)
\begin{table*}[t]
\caption{Fitted parameter values of the current model. For the details of the
resonance parameters, see the Appendix. For the contact amplitude, see
Eq.~(\ref{ampl_cont}). The entries in boldface are taken from Ref.~\cite{NOH06}
and they are not fit parameters. Here, it is assumed that  $\phi_a^{TL}=
\phi_a$ and $\phi_b^{TL}= \phi_b$, in addition to
$\alpha^{T}_{L}=\beta^T_{L}=\alpha$. }
\sisetup{detect-weight=true,detect-inline-weight=math}
\begin{tabular}{ c S[table-format=-2.3] S S S c || c S S r S S S S }
\hline
Y & {$ g_{\NLK}^{}$} & {\hspace{1mm} $\lambda_{N\Lambda K}^{}$} &
{\hspace{1mm}$g_{\XLK}^{}$} & {\hspace{1mm}$\lambda_{\XLK}^{}$} & \hspace{1mm}$\Lambda$ (MeV) &
 &  &  &  &  & \\ \hline
$\Lambda(1116)\frac12^+$ &  \bfseries -13.24 & \bfseries 1.0 & \bfseries 3.52
& \bfseries 1.0 & $900$
& & & & & \\
$\Sigma(1193)\frac 12^+$ & \bfseries 3.58 & \bfseries 1.0 &
\bfseries -13.26 & \bfseries 1.0 & $900$
& & & & & \\   \hline\hline
                & {$ g_{\NLK}^{}g_{\XLK}^{}$} & {--} & {--}
 & {--} & \hspace{1mm}$\Lambda$ (MeV) &
\hspace{2mm}$L$ & {\hspace{4.5mm}$a^0_{L}$\hspace{4.5mm}} & {\hspace{3mm}$a^1_{L}$\hspace{3mm}} & & {\hspace{3mm}$b^0_{L}$\hspace{3mm}}
& {\hspace{3mm}$b^1_{L}$\hspace{3mm}} & {\hspace{3mm}$\phi_a^{}$\hspace{3mm}} & {\hspace{3mm}$\phi_b^{}$\hspace{3mm}} \\ \hline
 $\Lambda(1890)\frac 32^+$ & 0.11 & & & & $900$
&\hspace{2mm} 0  & 0.28 & -1.19 & & & & &   \\
$\Sigma(1385)\frac 32^+$ & 18.76 &  &  &  & $900$
&\hspace{2mm} 1 & 3.23 & -4.84 & & -3.40 & 0.61 &  &  \\
$\Sigma(2030)\frac 72^+$ & 0.49 &  & & & $900$
&\hspace{2mm} 2 & 3.06  & 21.07 & & 9.40   &  -2.28 &  &  \\
$\Sigma(2250)\frac 52^-$ & -0.033 &  & & & $900$ &
\multicolumn{3}{ c }{\hspace{1.1cm}$\Lambda_S=1$ GeV} & \multicolumn{3}{c }{\hspace{6mm}$\alpha=3.60$}
& 0.22 & -0.16 \\  \hline
\end{tabular}
\label{tab:para_g_contact}
\end{table*}
%%%%%%%%%%%%%%%%%%%%%%%%%%%%%%%%%%%%%%%%%%%%%%

The resonances included in the present model calculations and the corresponding
resonance parameters are displayed in Table~\ref{tab:para_g_contact} as well as
the parameters of the phenomenological contact term, $M_c$. We do not give the
associated uncertainties here because they are not well constrained. In the
present calculation, resonances with $J \leq 7/2$ were considered. The masses
and the total widths of the resonances are taken to be those quoted in
PDG~\cite{PDG14} and are given in Table~\ref{tbl:hyperons}, except for the mass
of the $\Sigma(2250)$ resonance. Currently, the $\Sigma(2250)$ resonance is not
well established and has a three-star status~\cite{PDG14}. In fact, the PDG
does not even assign the spin-parity quantum numbers for this resonance. The
analyses of Ref.~\cite{DBRV72} provide two possible parameter sets, one with
$J^P=5/2^-$ at about $2270\pm50$ MeV and another with $J^P=9/2^-$ at about
$2210\pm30$ MeV. In the present work we have assumed $\Sigma(2250)$ to have
$J^P=5/2^-$ with the mass of 2265~MeV, the primary reason being that the total
cross section in $K^- + p \to K^+ + \Xi^-$ shows a small bump structure at
around 2300~MeV, which is well reproduced in our model with these parameter
values. For the corresponding width, we have adopted the value quoted in PDG as
shown in Table~\ref{tbl:hyperons}.

%%%%%%%%%%%%%%%%%%%%%%%%%%%%%%%%%%%%%%%%%%%%%%
%    Figure 2
\begin{figure*}[t]
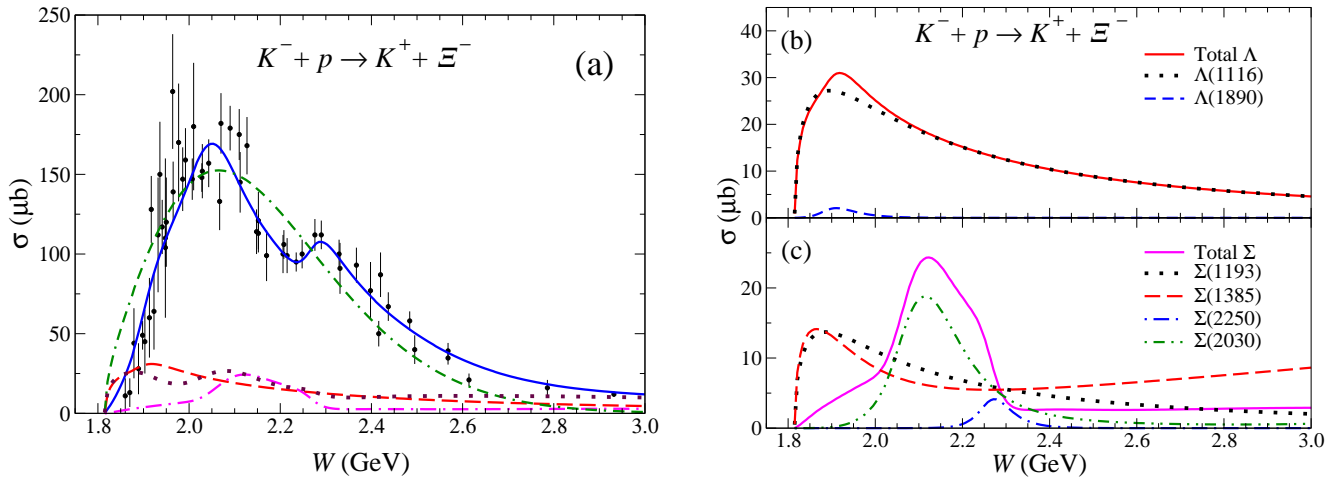
 \centering
\includegraphics[height=0.35\textwidth,clip=]{fig2a.eps} \qquad
\includegraphics[height=0.35\textwidth,clip=]{fig2b.eps}
\caption{(Color online)
Total cross section for the $K^-+p \rightarrow K^++\Xi^-$ reaction.
(a) The solid blue line represents the result of the full calculation of the present model.
The red dashed line shows the combined $\Lambda$ hyperons contribution.
The magenta dash-dotted line shows the combined $\Sigma$ hyperons contribution.
The brown dotted line shows the combined $\Lambda$ and $\Sigma$ hyperons contribution.
The green dash-dash-dotted line corresponds to the contact term.
(b)  The solid red line represents the combined $\Lambda$ hyperons contribution that
is the same as the red dashed line in (a).
The dotted and dashed lines show the $\Lambda(1116)$ and $\Lambda(1890)$
contributions, respectively.
(c) The solid magenta line represents the combined $\Sigma$ hyperons contribution that
is the same as the magenta dash-dotted line in (a).
The dotted, dashed, dot-dashed, and dot-dot-dashed lines show the contributions from the
$\Sigma(1193)$, $\Sigma(1385)$, $\Sigma(2250)$, and $\Sigma(2030)$, respectively.
The experimental data (black circles) are the digitized version as quoted in Ref.~\cite{SKL11}
from the original work of Refs.~\cite{CPSS64,BEHM66,BGLOR66,LRSY66,TS67,BMPT68,TFHL68,%
DBHM69,SABRE71,DBRV72,RBGP73,GGBP75,BGKS77}.
}
\label{xsc_channels_1A}
\end{figure*}
%%%%%%%%%%%%%%%%%%%%%%%%%%%%%%%%%%%%%%%%%%%%%%

All parameters of the present model calculation are determined as described
above and we now present the results obtained from our model. The overall fit
quality is quite good with $\chi^2/\text{d.o.f.} = 1.55$ or $\chi^2/N = 1.49$,
as displayed in Table~\ref{TABLE_res_chi2}. There, we also show the partial
$\chi$-squared values $\chi^2_i/N_i^{}$ evaluated for a given type of
observable specified by the index $i$ as explained in the caption of
Table~\ref{TABLE_res_chi2}. In Fig.~\ref{xsc_channels_1A} we show the results
for the total cross section in the charged $\Xi$ production reaction from the
proton target,  $K^-+p\rightarrow K^-+\Xi^-$, for the c.m. energies
up to $W=3$ GeV. Figure~\ref{xsc_channels_1A}(a) displays the total
contribution, which reproduces the data rather well. The dynamical content of
the present model is also shown in the same figure. We find that the contact
term rises quickly from threshold peaking at around $2.1$~GeV and falls off
slowly as energy increases. It dominates the cross section except for energies
very close to threshold and above $\sim 2.7$~GeV, where the hyperon resonance
contributions are comparable. The $\Lambda$ hyperons contribution is strongest
near threshold and falls off very slowly as energy increases. The $\Sigma$
hyperons contribution is relatively small over the entire energy range
considered, except in the interval of $ 2.0$--$2.3$~GeV, where it becomes
comparable to the $\Lambda$ contribution. Near threshold, there is a strong
destructive interference between the contact term and (mainly) the $\Lambda$
hyperons contribution. At higher energies, the data indicates an existence of a
bump structure at $W \sim 2.3~\mbox{GeV}$. Our model reproduces this feature
via delicate destructive and constructive interferences of the contact term and
the hyperon resonance contributions. We also mention that we have explored the
possibility of a much smaller contact amplitude contribution than shown in
Fig.~\ref{xsc_channels_1A}(a) considering various different sets of hyperon
resonances from Table~\ref{tbl:hyperons}; however, we were unable to find a
solution with a fit quality comparable to that of
Fig.~\ref{xsc_channels_1A}(a).

Figure \ref{xsc_channels_1A}(b) displays the individual $\Lambda$ hyperon
contributions. We see that the ground state $\Lambda(1116)$ is, by far, the
dominant contribution which is due to the tail of the corresponding $u$-channel
process. Analogously, the individual $\Sigma$ hyperon contributions are shown
in Fig.~\ref{xsc_channels_1A}(c). Here, the relatively small cross section near
threshold is due to the destructive interference between the $\Sigma(1192)$ and
$\Sigma(1385)$. The enhancement of the cross section in the energy interval of
$2.0$--$2.3$~GeV is mostly due to the $\Sigma(2030)$ resonance. The
$\Sigma(2250)$ leads to a little shoulder in the total $\Sigma$ contribution.
We note that any non-negligible contribution from the hyperons for energies
above $\sim 2.3$~GeV is due to the $u$-channel processes.

%%%%%%%%%%%%%%%%%%%%%%%%%%%%%%%%%%%%%%%%%%%%%%
%    Figure 3
\begin{figure*}[t]
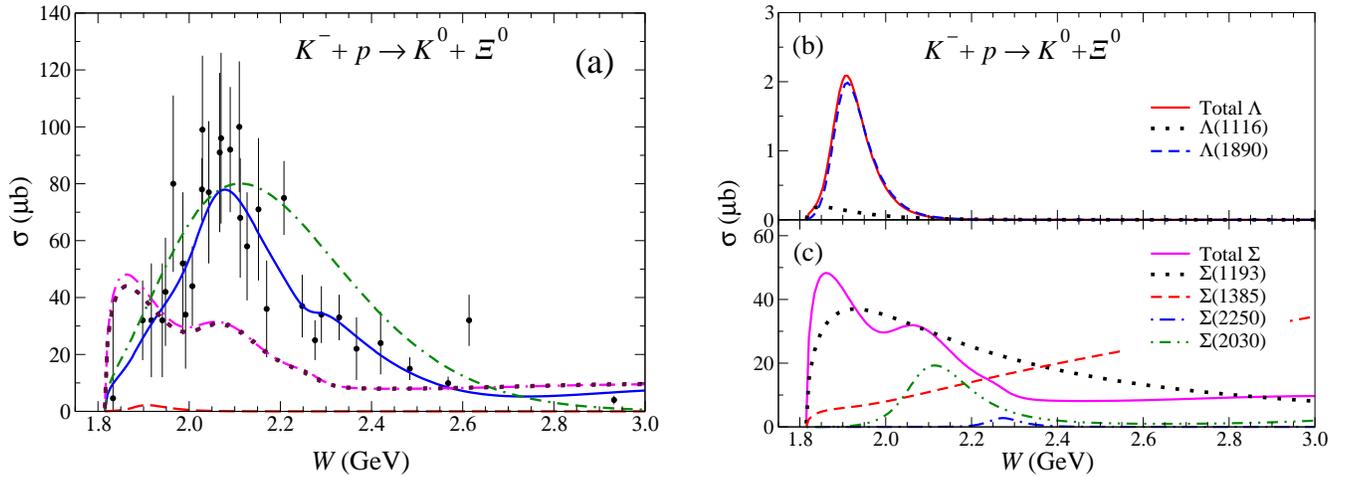
 \centering
\includegraphics[height=0.35\textwidth,clip=]{fig3a.eps}  \qquad
\includegraphics[height=0.35\textwidth,clip=]{fig3b.eps}
\caption{(Color online)
Same as Fig.~\ref{xsc_channels_1A} for the $K^-+p\rightarrow K^0+\Xi^0$ reaction.
The experimental data (black circles) are the digitized version as quoted in Ref.~\cite{SKL11} from
the original work of Refs.~\cite{BEHM66,DBHM69,SABRE71,DBRV72,CDJSE73,BGKS77}.
}
\label{xsc_channels_2A}
\end{figure*}
%%%%%%%%%%%%%%%%%%%%%%%%%%%%%%%%%%%%%%%%%%%%%%

In Fig.~\ref{xsc_channels_2A}, we show the total cross section results for the neutral $\Xi$ production process,
$K^-+p\rightarrow K^0+\Xi^0$.
Here, the data are of such poor quality that they impose much less constraint on the model parameters than the
corresponding data in the charged $\Xi^-$ production.
The resulting dynamical content shown in Fig.~\ref{xsc_channels_2A}(a) is similar to that for the charged $\Xi^-$
production discussed above, i.e., it is largely dominated by the contact term.
However, we see a quite different feature in the $\Lambda$ and $\Sigma$ resonance contributions as compared
to that for the charged $\Xi^-$ production [cf. Fig.~\ref{xsc_channels_1A}(a)].
One notable difference between the charged and neutral $\Xi$ production reactions considered here is that the
$u$-channel $\Lambda$ hyperon contribution is absent in the $\Xi^0$ production case.
Also, the relative contribution of the $\Sigma$ hyperons is much larger in the neutral $\Xi^0$ production than in the
charged $\Xi^-$ production, especially, in the near threshold region.

Figures~\ref{xsc_channels_2A}(b) and \ref{xsc_channels_2A}(c) show the
individual hyperon contributions. As mentioned before, due to the absence  of
the $u$-channel $\Lambda$ exchange in the neutral $\Xi^0$ production, the
$\Lambda(1116)$ contribution is insignificant, leading to a negligible
contribution of the $\Lambda$ hyperons. Due to the isospin factors here, the
$\Sigma(1192)$ and $\Sigma(1385)$ hyperons interfere constructively, especially
near the threshold. Recall that, for charged $\Xi^-$ production, these hyperons
interfere destructively [cf. Fig.~\ref{xsc_channels_1A}(c)].

%%%%%%%%%%%%%%%%%%%%%%%%%%%%%%%%%%%%%%%%%%%%%%
%    Figure 4
\begin{figure*}[t!]
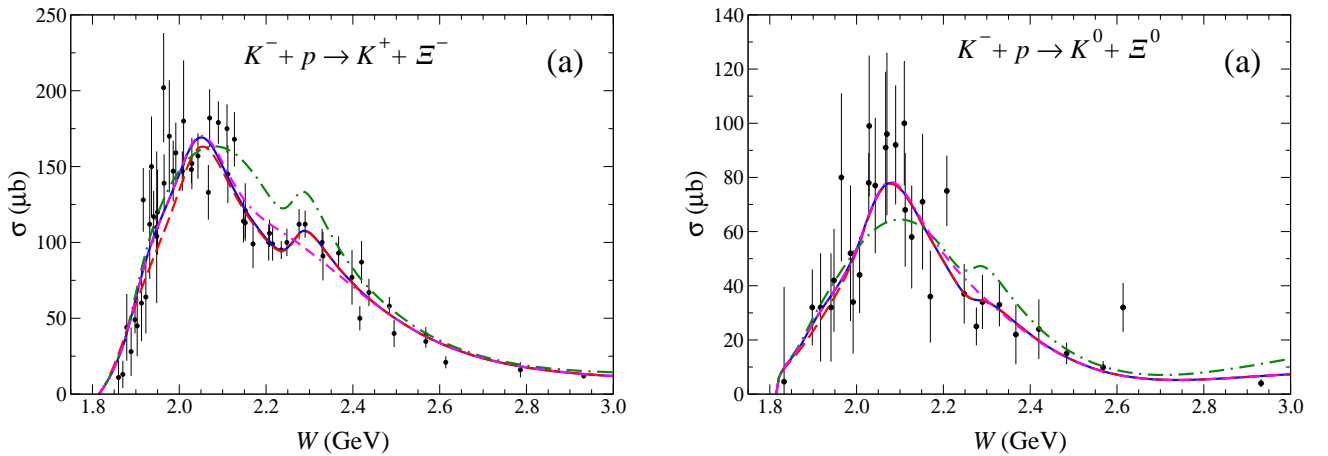
\centering
\includegraphics[width=0.46\textwidth,clip=]{fig4a.eps} \qquad
\includegraphics[width=0.46\textwidth,clip=]{fig4b.eps}
\caption{(Color online)
Total cross section results with individual resonances switched off
(a) for $K^-+p\rightarrow K^++\Xi^-$ and (b) for $K^-+p\rightarrow K^0+\Xi^0$.
The blue lines represent the full result shown in Figs.~\ref{xsc_channels_1A} and
\ref{xsc_channels_2A}.
The red dashed lines, which almost coincide with the blue lines represent the result with
$\Lambda(1890)$ switched off.
The green dash-dotted lines represent the result with $\Sigma(2030)$ switched off and the
magenta dash-dash-dotted lines represent the result with $\Sigma(2250) 5/2^-$ switched off.
}
\label{txsc_RR}
\end{figure*}
%%%%%%%%%%%%%%%%%%%%%%%%%%%%%%%%%%%%%%%%%%%%%%

In Fig.~\ref{txsc_RR} we illustrate the amount of the above-threshold resonance
contributions of the present model to the total cross sections. We do this by
comparing the full results shown in Figs.~\ref{xsc_channels_1A}(a) and
\ref{xsc_channels_2A}(a) to the result found by switching off one resonance at
a time. We see in Fig.~\ref{txsc_RR}(a) that the largest effect of
$\Sigma(2030)$ on the cross sections is in the range of $W \sim 2.0$ to
2.4~GeV. This resonance is clearly needed in our model to reproduce the data.
It also affects the recoil polarization as will be discussed later. It should
be mentioned that this resonance also helps to reproduce the measured
$K^+\Xi^-$ invariant mass distribution in $\gamma + p \to K^+ + K^+ +
\Xi^-$~\cite{MON11}, by filling in the valley between the two bumps in the
invariant mass distribution that would appear without it; such a feature
clearly is not observed in the data~\cite{CLAS07b}. The $\Lambda(1890)$ affects
the total cross section in the range of $W \sim 1.9$ to 2.1~GeV, and the
$\Sigma(2250)5/2^-$ contributes around $W \sim 2.2$~GeV, where it is needed to
reproduce the observed bump structure. A more accurate data set is clearly
needed for a more definitive answer about the roles of the $\Lambda(1890)$ and
$\Sigma(2250)$ resonances. Figure~\ref{txsc_RR}(b), for the neutral $\Xi^0$
production, also shows a similar feature observed in the $\Xi^-$ case for the
$\Sigma(2030)$ resonance. Here, the influence of the $\Sigma(2250)5/2^-$ is
smaller and  that of the $\Lambda(1890)$ is hardly seen. Recall that there is
no $u$-channel $\Lambda$ contribution in the neutral $\Xi^0$ production.

%%%%%%%%%%%%%%%%%%%%%%%%%%%%%%%%%%%%%%%%%%%%%%
%    Figure 5
\begin{figure*}[t!]
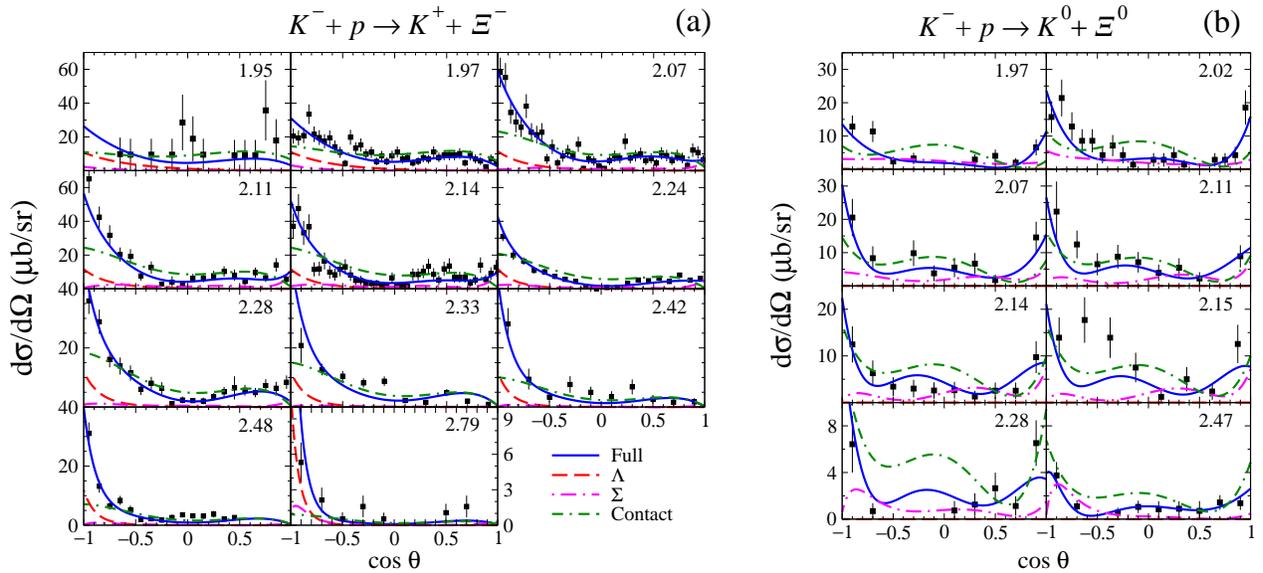
\centering
\includegraphics[height=0.42\textwidth,clip=]{fig5a.eps} \qquad
\includegraphics[height=0.42\textwidth,clip=]{fig5b.eps}
\caption{(Color online)
Kaon angular distributions in the center-of-mass frame (a) for $K^-+p\rightarrow K^++\Xi^-$ and (b) for
$K^-+p\rightarrow K^0+\Xi^0$.
The blue lines represent the full model results.
The red dashed lines show the combined $\Lambda$ hyperons contribution.
The magenta dash-dotted lines show the combined $\Sigma$ hyperons contribution.
The green dash-dash-dotted line corresponds to the contact term.
The numbers in the upper right corners correspond to the centroid total energy of the system $W$.
Note the different scales used.
The experimental data (black circles) are the digitized version as quoted in Ref.~\cite{SKL11} from
the original work of Refs.~\cite{BGLOR66,LRSY66,TS67,BMPT68,TFHL68,DBHM69}
for the $K^- + p \to K^+ + \Xi^-$ reaction and of Ref.~\cite{BEHM66,BMPT68,DBHM69,CDJSE73}
for the $K^- + p \to K^0 + \Xi^0$ reaction.
}
\label{dxsc_1}
\end{figure*}
%%%%%%%%%%%%%%%%%%%%%%%%%%%%%%%%%%%%%%%%%%%%%%

%%%%%%%%%%%%%%%%%%%%%%%%%%%%%%%%%%%%%%%%%%%%%%
%    Figure 6
\begin{figure*}[t]
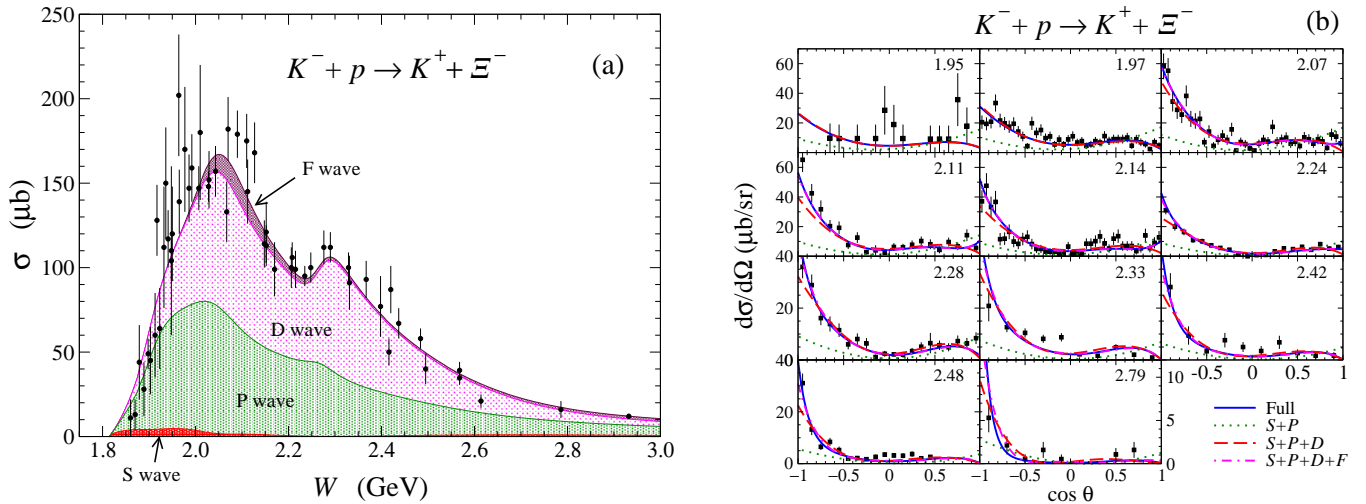
\centering
\includegraphics[height=0.37\textwidth,clip=]{fig6a.eps} \qquad
\includegraphics[height=0.37\textwidth,clip=]{fig6b.eps}
\caption{(Color online)
Partial wave decomposition of the total cross section and the angular distribution for $K^- + p \to K^+ + \Xi^-$.
(a) Total cross section sectioned by contributions from each partial wave $L$.
The red shaded area indicates the $S$-wave contribution, while the green area corresponds to the $P$-wave.
Magenta indicates the $D$-wave and maroon the $F$-wave.
(b) $K^+$ angular distribution:  the solid blue lines are the full results, while the dotted green lines represent
the sum of $S+P$ waves, the red dashed lines represent the $S+P+D$ waves and the dash-dotted
magenta lines represent the $S+P+D+F$ waves.
For lower energies, the $S+P+D$ waves already saturate the full cross section results so that the
$F$- and higher-wave contributions cannot be distinguished from the full result.
}
\label{results_PW1}
\end{figure*}
%%%%%%%%%%%%%%%%%%%%%%%%%%%%%%%%%%%%%%%%%%%%%%

\begin{figure*}[t]
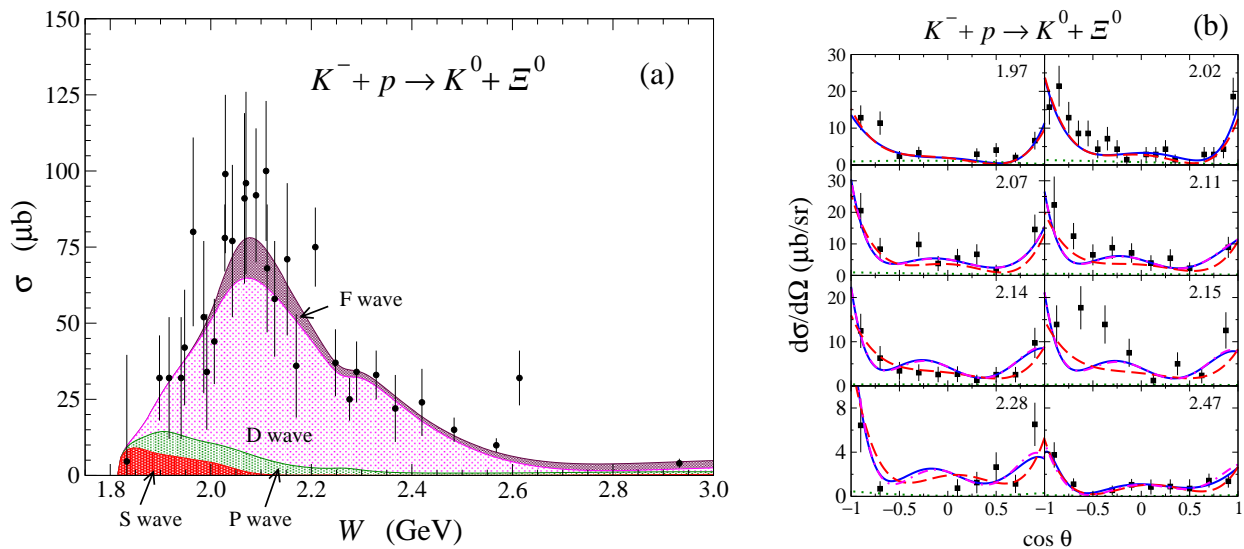
\centering
\includegraphics[height=0.4\textwidth,clip=]{fig8a.eps} \qquad
\includegraphics[height=0.4\textwidth,clip=]{fig8b.eps}
\caption{(Color online)
Same as in Fig.~\ref{results_PW1} but for $K^- + p \to K^0 + \Xi^0$. }
\label{results_PW2}
\end{figure*}

The results for differential cross sections in both $K^- + p \to K^+ + \Xi^-$ and $K^- + p \to K^0 + \Xi^0$ are
shown in Figs.~\ref{dxsc_1}(a) and \ref{dxsc_1}(b), respectively, in the energy domain up to $W =2.8$~GeV
for the former and up to $W=2.5$~GeV for the latter reaction.
Overall, the model reproduces the data quite well.
%There seem to be some discrepancies
%in the energy range $W=2.33$ to $2.48$ GeV in the charged $\Xi^-$
%production. Our model underpredicts the yield around $\cos\theta=0$.
As in the total cross sections, the data for the neutral $\Xi^0$ production are fewer and less accurate than for the
charged $\Xi^-$ production.
In particular, the $\Xi^0$ production data at $W=2.15$~GeV seems incompatible with those at nearby energies,
and the present model is unable to reproduce the observed shape at backward angles.
It is clear from Figs.~\ref{dxsc_1}(a) and \ref{dxsc_1}(b) that the charged channel shows a backward peaked
angular distributions, while the neutral channel shows enhancement for both backward and forward scattering angles
(more symmetric around $\cos\theta=0$) for all but perhaps the highest energies.
In the charged $\Xi^-$ production, both the resonance and contact amplitude contributions are backward angle
peaked and, as the energy increases, they get smaller and smaller at forward angles.
In $\Xi^0$ production, both the $\Sigma$ resonance and contact amplitude contributions also exhibit an
enhancement for forward angles.
Note that the $\Lambda$ resonance contribution here is negligible due to the absence of the $u$-channel process.
The interference pattern in the forward angular region depends on energy.
At lower energies the interference is constructive and it becomes destructive at higher energies.
The behavior of the angular distributions in terms of the partial waves will be discussed later in connection with the
results of Figs.~\ref{results_PW1}(b) and \ref{results_PW2}(b).

%%%%%%%%%%%%%%%%%%%%%%%%%%%%%%%%%%%%%%%%%%%%%%
%    Figure 7
\begin{figure}[t]\centering
\includegraphics[width=0.45\textwidth,clip=]{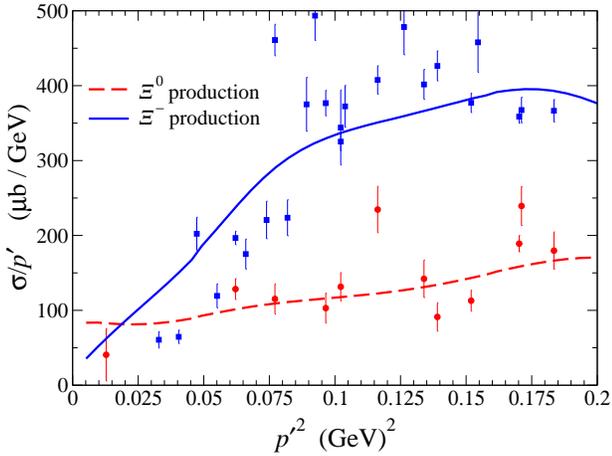} \qquad
\caption{(Color online)
Ratio of the measured total cross section $\sigma$ and the final state $K\Xi$ relative momentum
$p'$ as a function of $p'^2$.
The blue square data correspond to $K^- + p \to K^+ + \Xi^-$, while the red circle data to
$K^- + p \to K^0+\Xi^0$.
The blue solid  and red dashed curves are the present model results corresponding to
$K^- + p \to K^+ + \Xi^-$ and $K^- + p \to K^0 + \Xi^0$, respectively.
}
\label{psigma}
\end{figure}
%%%%%%%%%%%%%%%%%%%%%%%%%%%%%%%%%%%%%%%%%%%%%%

%%%%%%%%%%%%%%%%%%%%%%%%%%%%%%%%%%%%%%%%%%%%%%
%    Figure 8
%%%%%%%%%%%%%%%%%%%%%%%%%%%%%%%%%%%%%%%%%%%%%%

The partial-wave content of the cross sections for the charged $\Xi^-$
production process arising from the present model is shown in
Figs.~\ref{results_PW1}(a) and \ref{results_PW1}(b). As can be seen in
Fig.~\ref{results_PW1}(a), the total cross section is dominated by the $P$ and
$D$ waves in almost the entire energy range considered, even at energies very
close to threshold where one sees a strongly rising $P$-wave contribution. The
$S$-wave contribution is very small. This peculiar feature is caused by the
ground state $\Lambda(1116)$, whose contribution cancels to a large extent the
otherwise dominant $S$-wave contribution close to threshold, in addition to
enhancing the $P$-wave contribution. One way of probing the $S$-wave content
close to threshold in a model-independent manner is to look at the quantity
$\sigma / p'$ as a function of $p'^2$, where $p'$ is the relative momentum of
the final $K\Xi$ state. The reason being that, for hard processes, the
partial-wave reaction amplitude behaves basically like $p'^L$ for a given
orbital angular momentum $L$ as mentioned in Sec.~\ref{sec:model}. This leads
to
\begin{equation}
\frac{\sigma}{p'} = c_0^{} + c_1^{} p'^2 + c_2^{} p'^4 + \dots \, ,
\label{eq:psigma}
\end{equation}
with expansion constants $c_L^{}$.
Figure~\ref{psigma} illustrates this point.
Although the existing experimental data are of poor quality, they reveal the general features just mentioned.
In particular, for the charged $\Xi^-$ production process, the data indicate a linear behavior of $\sigma / p'$ close
to threshold implying a strong $P$-wave contribution.
Our present model result is consistent with this behavior.
It is also consistent with the observation made in Ref.~\cite{KNLS14} that the low-energy
behavior of the total cross sections in the $\pi N$, $\eta N$ and $K\Xi$
channels does not seem to follow the usual $S$-wave dominance.

The corresponding results for the neutral $\Xi^0$ production are also shown in
Fig.~\ref{psigma}. There, the scattered data are consistent with $S$-wave
dominance, a feature exhibited by our model as well [see also
Fig.~\ref{results_PW2}(a)]. In Fig.~\ref{results_PW1}(a), we also see a small
$F$-wave contribution above $W \sim 2.0$~GeV that helps saturate the total
cross section. Note that since our contact term includes partial waves only up
to $L\le2$, the $F$-wave contribution is entirely due to the hyperon
resonances. The enhancement of the $D$-wave contribution around $W=2.3$ GeV as
well as the little shoulder in the $P$-wave contribution are due to the
$\Sigma(2250)$ hyperon. Of course, the partial-wave contributions are mainly
constrained by the differential cross section and they are shown in
Fig.~\ref{results_PW1}(b). As mentioned before, the shape of the angular
distribution is backward-angle peaked and the cross sections are very small at
forward angles. This behavior is a direct consequence of the very significant
interference between the $P$ and $D$ waves. This can be seen by expanding the
cross section in partial waves. Considering the partial waves through $L=2$ and
following Ref.~\cite{JOHN14}, the differential cross section may be expressed
as
\begin{align}
\frac{d\sigma}{d\Omega} & =   \left|\alpha_{02}^{} \right|^2 + \left[ \left|\alpha_1^{}\right|^2
+ 2\,\mbox{Re}\left( \alpha_{02}^{} \tilde\alpha_2^*\right) \right] \cos^2\theta
\nonumber\\
 & \mbox{}\quad
+  \left| \tilde\alpha_2^{} \right|^2 \cos^4\theta
+   \left( | \beta_1^{} |^2 +  | \tilde{\beta}_2^{} |^2 \cos^2\theta \right) \sin^2\theta
\nonumber \\
 &\mbox{}\quad
+ 2 \, \mbox{Re} \left[ \alpha_{02}^{}\alpha_1^* + \alpha_1^{} \tilde\alpha_2^*\cos^2\theta +
\beta_1^{}\tilde{\beta}_2^*\sin^2\theta \right]\cos\theta~,
\label{PW-dxsc}
\end{align}
where the coefficients $\alpha_{L}^{}$ ($\beta_{L}^{}$) provide a linear combination of the partial-wave matrix
elements corresponding to the spin-non-flip (spin-flip) process with a given orbital angular momentum $L$
[see Eq.~(\ref{eq:PW})].
Here, $\alpha_{02}^{} \equiv \alpha_0^{} - \frac13 \tilde\alpha_2^{}$,
$\tilde\alpha_2^{}\equiv\frac23\alpha_2^{}$, and $\tilde{\beta}_2^{} \equiv 3\beta_2^{}$.
In the above equation, the last term on the right-hand side involving  an interference between the $P$ and $D$
waves is an odd function in $\cos\theta$, while the the first term in square brackets is an even function.
These two terms cancel to a large extent at forward angles while at backward angles they add up.
Note that these partial waves are comparable in strength as shown in Fig.~\ref{results_PW1}(a) so that their
interference term leading to an odd function part can largely cancel the even term at forward angles.

Figures~\ref{results_PW2}(a) and \ref{results_PW2}(b) display the partial wave content in the cross sections for
the neutral $\Xi^0$ production process.
In contrast to the charged $\Xi^-$ production, here one sees that the largest contribution to the total cross section
is the $D$-wave, and  the $P$-wave is largely suppressed, which is a direct consequence of the shape of
the observed angular distribution whose partial wave contributions are shown in Fig.~\ref{results_PW2}(b).
There, compared to that for charged $\Xi^-$, one sees a more symmetric angular shape about $\cos\theta=0$
that is dominated by the $D$-wave.
The present model reproduces the observed behavior of the $K^0$ angular distribution by suppressing the
$P$-wave contribution as can be easily understood from Eq.~(\ref{PW-dxsc}).
The rather drastic suppression of the $P$ wave can be better seen in Fig.~\ref{results_PW2}(a).
For energies very close to threshold, the cross section is dominated by the $S$-wave as seen also in
Fig.~\ref{results_PW2}(a).

%%%%%%%%%%%%%%%%%%%%%%%%%%%%%%%%%%%%%%%%%%%%%%
%    Figure 9
\begin{figure}[h!]\centering
\includegraphics[width=0.45\textwidth,clip=]{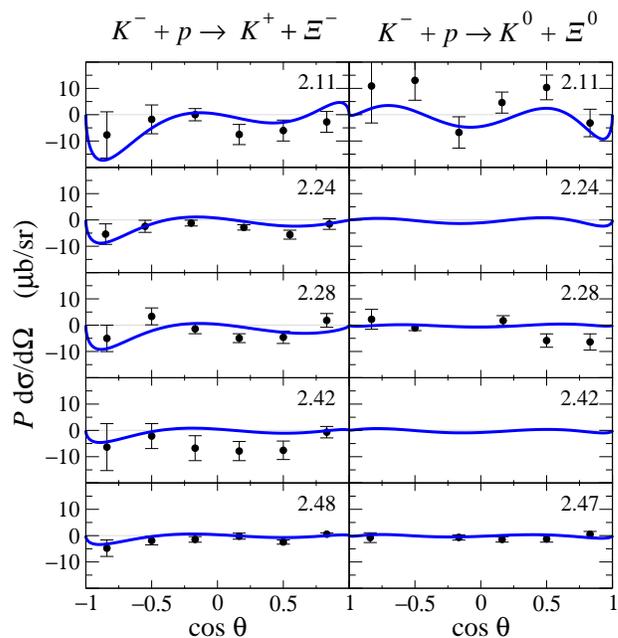}
\caption{(Color online) The recoil asymmetry
multiplied by the cross section,
$P \frac{d\sigma}{d\Omega}$,
for both the $K^-+p\rightarrow K^++\Xi^-$ and $K^-+p\rightarrow K^0+\Xi^0$
reactions.
The blue solid lines represent the full results of the current model.
Data are from Refs.~\cite{TS67,DBHM69}.
}
\label{P_y_12_phase}
\end{figure}
%%%%%%%%%%%%%%%%%%%%%%%%%%%%%%%%%%%%%%%%%%%%%%

The results for the recoil polarization asymmetry multiplied by the cross
section are shown in Fig.~\ref{P_y_12_phase} in the energy interval of $W=2.1$
to 2.5 GeV. Overall, we reproduce the data reasonably well. We also find that
the results shown at $W= 2.11$ GeV are still significantly affected by
$\Sigma(2030)$. This corroborates the findings of Ref.~\cite{SKL11}. We recall
that the recoil asymmetry is proportional to the imaginary part of the product
of the non-spin-flip matrix element ($M_{ss}$) with the complex conjugate of
the spin-flip matrix element ($M_{s's}$ with $s' \ne s$)~\cite{JOHN14}, so that
it vanishes identically unless these matrix elements are
such that their product has a non-vanishing imaginary part.
We can therefore
expect the recoil polarization to be sensitive to the complex nature of the
reaction amplitude, in particular, to the phenomenological contact amplitude,
$M_c$, introduced in the present model. Indeed, if one forces the coupling
strength parameters, $g_1^{LT}$ and $g_2^{LT}$ in Eq.~(\ref{ampl_cont}), to be
pure real during the fitting procedure, the $\chi^2_P/N_P$ deteriorates, e.g.,
from 1.89 to 2.26 for the $K^- + p \to K^+ + \Xi^-$ reaction, although the
quality of fit for cross sections is nearly unchanged.

%%%%%%%%%%%%%%%%%%%%%%%%%%%%%%%%%%%%%%%%%%%%%%
%    Figure 10
\begin{figure*}[t]
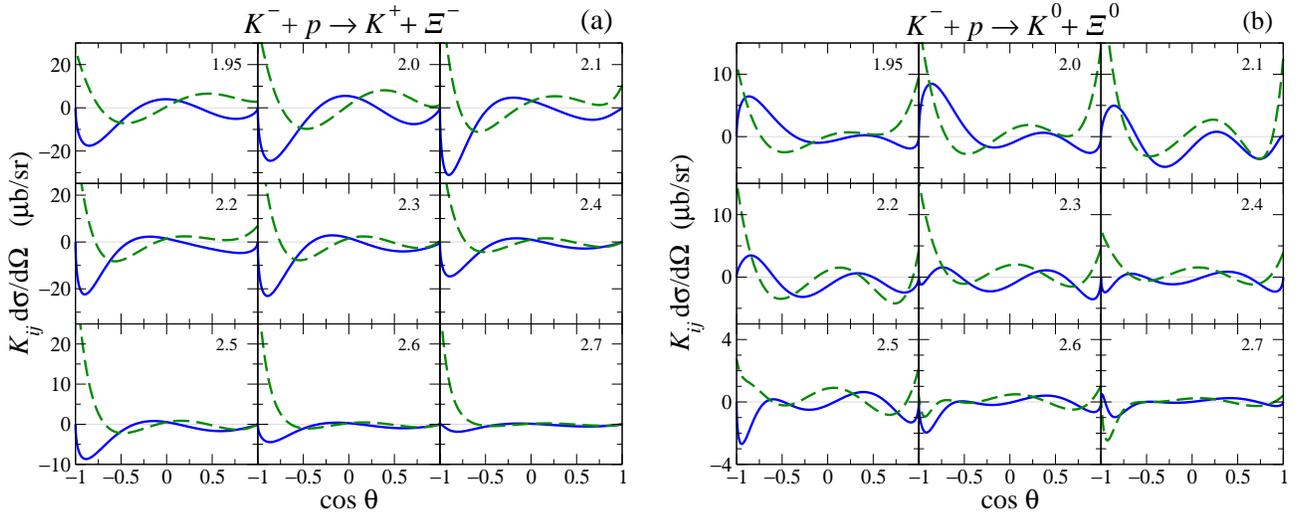
\centering
\includegraphics[width=0.46\textwidth,clip=]{fig10a.eps} \ \ \ \
\includegraphics[width=0.46\textwidth,clip=]{fig10b.eps}
\caption{(Color online)
Target-recoil asymmetries $K_{xx}$ (green dashed curves) and $K_{xz}$ (blue solid curves)
as defined in Ref.~\cite{JOHN14} multiplied by the cross sections
for the reactions
(a) $K^- + p \to K^+ + \Xi^-$ and (b) $K^- + p \to K^0 + \Xi^0$.
The numbers in the upper right corners represent the total energy of the system
$W$ in units of GeV.
 }
\label{POL_A1}
\end{figure*}
%%%%%%%%%%%%%%%%%%%%%%%%%%%%%%%%%%%%%%%%%%%%%%

In Fig.~\ref{POL_A1}, we show the present model predictions for the target-beam
asymmetries, $K_{xx}$ and $K_{xz}$, multiplied by the unpolarized cross
section, i.e., $\frac{d\sigma}{d\Omega}K_{xx}$ and
$\frac{d\sigma}{d\Omega}K_{xz}$ for both the charged $\Xi^-$ and neutral
$\Xi^0$ production processes. These observables are related to the spin-rotation parameter
$\beta$~\cite{Wolfen56} by $\tan\beta = -  K_{xz} / K_{xx}$. Note that these
target-recoil asymmetries, together with $K_{yy}$, are the only three
independent double-spin observables in the reaction of Eq.~(\ref{reac}) as
discussed in Ref.~\cite{JOHN14}. Indeed, the only two other non-vanishing
target-recoil asymmetries are related by $K_{zz} = K_{xx}$
and $K_{zx} = - K_{xz}$.%
\footnote{Note that the symmetry of the reaction leads to $K_{yy} = \pi_\Xi^{}$
   independent on the scattering angle $\theta$~\cite{NOH12,JOHN14}. Here,
   $\pi_\Xi^{}$ stands for the parity of the produced $\Xi$ which is taken to be
   $\pi_\Xi^{} = +1$ for the ground state $\Xi$. Also, $K_{xx} =
   K_{zz}\big|_{\cos\theta=\pm1}=\pi_\Xi^{}$. The target asymmetry is identical to
   the recoil asymmetry in the present reaction. Therefore, we exhaust all the
   \textit{independent} observables available in the reaction processes considered
   here.}
We mention that $\frac{d\sigma}{d\Omega}K_{xx}$ is proportional to the
difference of the magnitude squared of the spin-non-flip and spin-flip matrix
elements, while $\frac{d\sigma}{d\Omega}K_{xz}$ is proportional to the real
part of the product of the spin-non-flip matrix element with the complex
conjugate of the spin-flip matrix element. Therefore, unlike the recoil
asymmetry, these spin observables do not vanish even if the reaction amplitude
is pure real or pure imaginary. This means that they are, like the cross
section, much less sensitive to the complex nature of the phenomenological
contact amplitude.

To gain some insight into the angular dependence exhibited by these target-recoil asymmetries in
Fig.~\ref{POL_A1}, we express them in terms of partial waves with $L \le 2$, which gives
\begin{subequations}\label{PW-BT}
\begin{align}
\frac{d\sigma}{d\Omega}K_{xx}
& =
\left|\alpha_{02}^{}\right|^2 + \left[\left|\alpha_1^{}\right|^2
+ 2 \, \mbox{Re} \left(\alpha_{02}^{} \tilde\alpha_2^*\right)\right] \cos^2\theta
\nonumber \\ & \mbox{}\quad
+ \left|\tilde\alpha_2^{}\right|^2\cos^4\theta
- \left( |\beta_1^{} |^2 + |\tilde{\beta}_2^{} |^2 \cos^2\theta \right) \sin^2\theta
\nonumber \\ & \mbox{}\quad
+ 2 \, \mbox{Re} \Big[ \alpha_{02}^{}\alpha_1^* + \alpha_1^{} \tilde\alpha_2^* \cos^2\theta
\nonumber\\
&\mbox{}\hspace{26mm}
- \beta_1^{}\tilde{\beta}_2^* \sin^2\theta \Big] \cos\theta ~ ,
\\
\frac{d\sigma}{d\Omega}K_{xz} & =  2\, \mbox{Re} \left[ \alpha_{02}^{}\beta_1^* +
\left(\alpha_1^{}\tilde{\beta}_2^* + \tilde\alpha_2^{}\beta_1^*\right)\cos^2\theta \right.
\nonumber \\ & \mbox{}\quad
+ \left. \left(\alpha_{02}^{}\tilde{\beta}_2^* + \alpha_1^{}\beta_1^*\right) \cos\theta +
\tilde\alpha_2^{} \tilde{\beta}_2^* \cos^3\theta \right] \sin\theta ~.
\end{align}
\end{subequations}
Note that the only difference between $\frac{d\sigma}{d\Omega}K_{xx}$ given
above and differential cross section given by Eq.~(\ref{PW-dxsc}) is the sign
change of the terms involving $\beta_{L}^{}$. These terms are, however,
proportional to $\sin^2\theta$. Therefore, this spin observable behaves like
the differential cross section at very forward and backward angles, where
$\sin^2\theta \ll 1$. At $\cos\theta =0$, the difference is due to the term of
$\pm |\beta_1|^2$, which is a $P$-wave contribution in the spin-flip amplitude.
Now, if we ignore the $P$-wave contribution --- which is relatively very small
in the neutral $\Xi^0$ production over the nearly entire energy region
considered as seen in Fig.~\ref{results_PW2}(a) --- it is immediate to see that
Eq.~(\ref{PW-BT}a) involves only terms that are symmetric about $\cos\theta=0$.
We see in Fig.~\ref{POL_A1}(b) that $\frac{d\sigma}{d\Omega}K_{xx}$ exhibits
roughly this symmetry.

For $\frac{d\sigma}{d\Omega}K_{xz}$, Eq.~(\ref{PW-BT}b) reveals a rather complicated angular
dependence in general, and no particular feature is apparent in the results shown in
Fig.~\ref{POL_A1}, especially for the charged $\Xi^-$ production process.
Neglecting the $P$-wave contribution,  Eq.~(\ref{PW-BT}b) reduces to
$\frac{d\sigma}{d\Omega}K_{xz} = \mbox{Re} \left[\left(\alpha_{02}^{} +
\tilde\alpha_2^{}\cos^2\theta\right)\tilde{\beta}_2^* \right] \sin2\theta$, which is roughly the angular
dependence exhibited in Fig.~\ref{POL_A1}(b).

%%%%%%%%%%%%%%%%%%%%%%%%%%%%%%%%%%%%%%%%%%%%%%
%    Figure 11
\begin{figure*}[t]
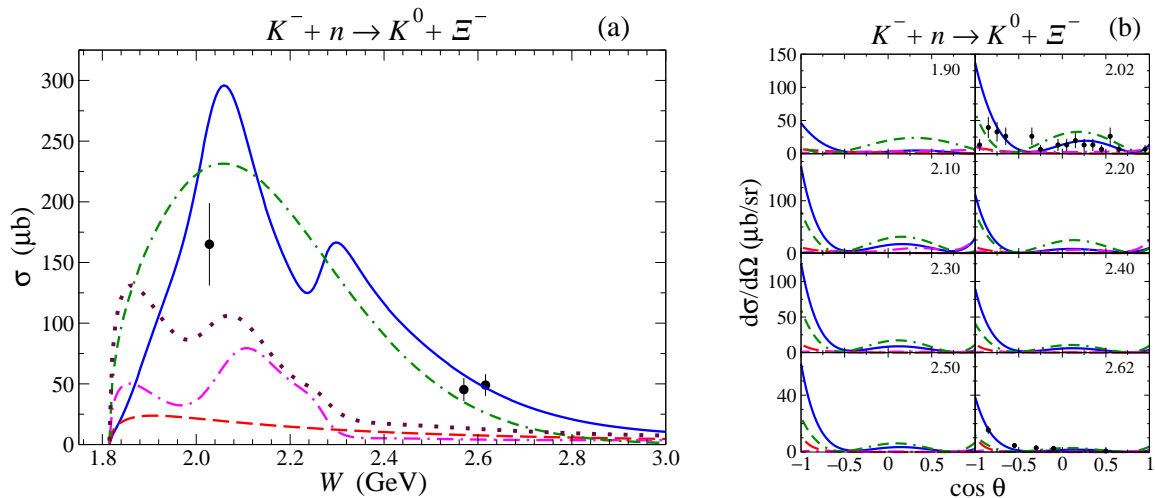
 \centering
\includegraphics[height=0.365\textwidth,clip=]{fig11a.eps}  \qquad
\includegraphics[height=0.365\textwidth,clip=]{fig11b.eps}
\caption{(Color online)
Same as Figs.~\ref{xsc_channels_1A}(a) and \ref{dxsc_1} for the $K^-+n\rightarrow K^0+\Xi^-$ reaction.
The experimental data are from Refs.~\cite{BEHM66,SABRE71}.}
\label{xsc_channels_4A}
\end{figure*}
%%%%%%%%%%%%%%%%%%%%%%%%%%%%%%%%%%%%%%%%%%%%%%

The present model predictions for the $K^- + n \to K^0 + \Xi^-$ reaction are
shown in Fig.~\ref{xsc_channels_4A}. Here, the experimental data are extremely
scarce, and they were not included in the present fitting procedure.
Nevertheless, the current model is seen to predict those few data quite
reasonably. Both the total and differential cross sections exhibit a very
similar feature to those of the $K^- + p \to K^+ + \Xi^-$ reaction with a
noticeable small enhancement in the differential cross sections as seen in
Fig.~\ref{xsc_channels_4A}(b) for forward angles near $\cos\theta=0$ in $K^- +
n \to K^0 + \Xi^-$. We see, however, some bigger differences in the individual
amplitude contributions, more clearly seen in the total cross sections that are
given in Fig.~\ref{xsc_channels_4A}(a). There, the $\Sigma$ hyperon
contribution is larger than the $\Lambda$ contribution over the entire energy
region up to $W \sim 2.3$ GeV, in particular, at low energies near threshold.
This is due to the absence of the strong destructive interference between the
$\Sigma(1385)$ and $\Sigma(1192)$ (not shown), since the latter hyperon
contribution is suppressed to a large extent compared to the case of $K^- + p
\to K^+ + \Xi^-$. Moreover, there is a constructive interference with the
$\Lambda$ hyperon, which makes the sum of the hyperons contribution relatively
large in the low energy region.

\begin{figure*}[t]
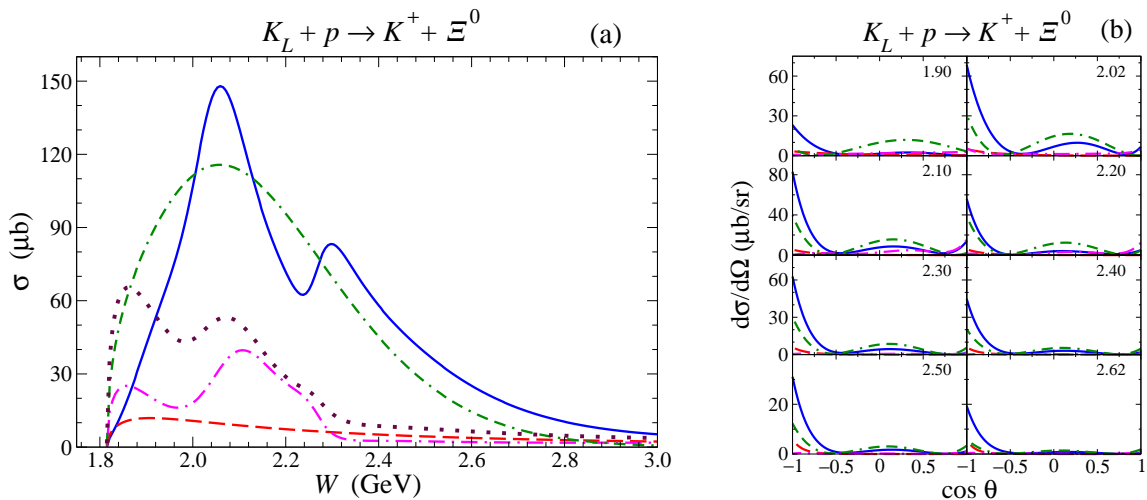
 \centering
\includegraphics[height=0.365\textwidth,clip=]{fig12a.eps}  \qquad
\includegraphics[height=0.365\textwidth,clip=]{fig12b.eps}
\caption{(Color online)
Same as Figs.~\ref{xsc_channels_1A}(a) and \ref{dxsc_1} for the $K_L+p\rightarrow K^++\Xi^0$ reaction.}
\label{fig:Klong}
\end{figure*}

For completeness, we also show in Fig.~\ref{fig:Klong} results for the $K_L + p
\to K^+ + \Xi^0$ reaction. Within the present model, the cross sections for
this process simply differ by a factor of 1/2 from those shown in
Fig.~\ref{xsc_channels_4A} for the $K^- + n \to K^0 + \Xi^-$ reaction. We show
the $K_L$ results here because the creation of a high-intensity $K_L$ beam
currently being contemplated~\cite{A15} may open up an entire new and exciting
field of hyperon spectroscopy.

%%%%%%%%%%%%%%%%%%%%%%%%%%%%%%%%%%%%%%%%%%%%%%
%    Figure 11
%%%%%%%%%%%%%%%%%%%%%%%%%%%%%%%%%%%%%%%%%%%%%%

%%%%%%%%%%%%%%%%%%%%%%%%%%%%%%%%%%%%%%%%%%%

\section{Conclusion} \label{sec:conclusion}

In this work we have presented our analysis on the reaction of $K^- + N \to K +
\Xi$ within an effective Lagrangian approach that includes a phenomenological
contact term to account for the final-state-interaction rescattering
contribution of the reaction amplitude in the Bethe-Salpeter equation and for
other possible (short-range) dynamics that are not explicitly taken into
account in the model. By introducing this phenomenological contact term, we
avoid the problems found in the usual implementations of tree-level effective
Lagrangian approaches that need to phenomenologically suppress $u$-channel
contributions dominating the high-energy behavior~\cite{SKL11,SST11}. In
addition to the ground states $\Lambda(1116)$ and $\Sigma(1193)$, the present
model also includes the $\Lambda(1890)$, $\Sigma(1385)$, $\Sigma(2030)$, and
$\Sigma(2250)$ resonance contributions.

The available total and differential cross sections, as well as the recoil
asymmetry data, in both the $K^- + p \to K^+ + \Xi^-$ and $K^- + p \to K^0 +
\Xi^0$ processes are well reproduced by the present model. We have found that
the above-threshold resonances $\Lambda(1890)$, $\Sigma(2030)$, and
$\Sigma(2250)$ are required to achieve a good fit quality for the data. Among
them, the $\Sigma(2030)$ resonance is the most critical one. This resonance
affects not only the cross sections but also the recoil asymmetry. In addition,
it also brings a model calculation of Ref.~\cite{MON11} into an agreement with
the observed $K^+\Xi^-$ invariant mass distribution in $\Xi$
photoproduction~\cite{CLAS07b}. The $\Lambda(1890)$ is also required to improve
the fit quality in the present model, especially in the energy dependence of
the total cross sections of the charged $\Xi^-$ production around $W=1.9$ GeV.
The total cross section data in the charged $\Xi^-$ production seems to
indicate a bump structure at around $W=2.3$ GeV, which is accounted for by the
$\Sigma(2250)$ resonance with $J^P=5/2-$ and a mass of 2265~MeV in the present
model. More accurate data are required before a more definitive answer can be
provided for the role of these two resonances. In this regard, the
multi-strangess hyperon production programs using an intense antikaon beam at
J-PARC is of particular relevance in providing the much needed higher-precision
data for the present reaction.

The present analysis also reveals a peculiar behavior of the total cross
section data in the threshold-energy region of the $K^+\Xi^-$ production
channel, where the higher partial-waves ($P$ and $D$) dominate instead of the
usual $S$-wave (cf.\ Fig.~\ref{psigma}). If this behavior of the cross section
data is corroborated in future experiments, it will cast serious doubts on the
validity of model calculations that neglect higher partial-waves even for
energies very close to threshold. This peculiar low-energy behavior of the
total cross section in $\bar{K}$-induced reaction seems to be present also in
the $\pi N$ and $\eta N$ production channels, in addition to the $K\Xi$ channel
\cite{KNLS14}.

Apart from the recoil asymmetry, we have also predicted the target-recoil
asymmetries for which there are currently no experimental data. In contrast to
the recoil polarization --- which are small --- these observables are quite
sizable and may help impose more stringent constraints on the model parameters.
In principle, one requires the four independent observables calculated here to
completely determine the reaction amplitude~\cite{JOHN14}. Of course,
measurements of the spin obervables, in particular, are challenging
experimentally by any standard, but one may exploit the self-analyzing nature
of the produced hyperon to help extract these
observables~\cite{JOHN14,CLAS14-a}. For the target-recoil asymmetry
measurements, one requires a polarized target in addition to spin measurements
of the produced $\Xi$. Polarized targets available at some of the world's major
laboratories combined with the availability of intense beams make measuring
these spin observables no longer out of reach. In fact, various single- and
double-polarization observables in photoproduction reactions are currently
being measured at  major facilities such as JLab, ELSA, and MAMI, aiming at
so-called complete experiment sets in order to model-independently determine
the photoproduction amplitudes.

While it may perhaps not be entirely clear which role any particular resonance
plays for the $K^- + N \to K + \Xi$ reaction, the present and other
calculations based on effective Lagrangians~\cite{SKL11,SST11}, and also the
unitarized chiral perturbation approach~\cite{MFR14}, seem to agree that some
$S=-1$ hyperon resonances seem to be required to reproduce the existing data.
To pin down the role of a particular resonance among them requires more precise
and complete data, in addition to more complete theoretical models. In any
case, the present reaction is very well suited for studying $S=-1$ hyperon
resonances.

Finally, the present work is our first step toward building a more complete
reaction theory to help analyze the data and extract the properties of $\Xi$
resonances in future experimental efforts in $\Xi$ baryon spectroscopy. This is
a complementary work to that of a model-independent analysis performed recently
by the same authors~\cite{JOHN14} and will also help in analyzing the data to
understand the production mechanisms of $\Xi$ baryons.

%%%%%%%%%%%%%%%%%%%%%%%%%%%%%%%%%%%%%%%%%%%

\acknowledgments

\newblock
We are grateful to D. A. Sharov for providing us with the digitized version of the data used 
in the present work.
\newblock
This work was partially supported by the National Research Foundation of Korea funded by 
the Korean Government (Grant No.~NRF-2011-220-C00011).
\newblock
The work of Y.O. was also supported in part by the Ministry of Science, ICT, and Future Planning 
(MSIP) and the National Research Foundation of Korea under Grant
No.\ NRF-2013K1A3A7A06056592 (Center for Korean J-PARC Users).
\newblock
The work of K.N. was also supported in part by the FFE-COSY Grant No. 41788390.

%%%%%%%%%%%%%%%%%%%%%%%%%%%%%%%%%%%%%%%%%%%

\appendix*
\section{}

In this Appendix, we give the effective Lagrangians and phenomenological
dressed baryon propagators from which the $s$- and $u$-channel amplitudes, $M_s$
and $M_u$ discussed in Sec.~\ref{sec:model}, are constructed. We follow
Refs.~\cite{NH04,NH05,NOH06,NOH08,MON11} and consider not only the spin-1/2
ground state $\Lambda$ and $\Sigma$ but also their respective excited states
with spin up to 7/2. In the following we use the notations for the iso-doublet
fields
\begin{align}
N &=
\left(\begin{array}{c}
p \\
n
\end{array}\right)~,
&
\Xi &=
\left(\begin{array}{c}
\Xi^0 \\
- \Xi^-
\end{array}\right)~,
\nonumber \\[-1ex]
\\[-1ex]
K &=
\left(\begin{array}{c}
K^+ \\
K^0
\end{array}\right)~,
&
K_c &=
\left(\begin{array}{c}
\bar{K}^0 \\
- K^-
\end{array}\right)~,
\nonumber
\end{align}
and for the iso-triplet fields
\begin{equation}
\bm{\Sigma} =
\left(\begin{array}{c}
\Sigma^+ \\
\Sigma^0 \\
\Sigma^-
\end{array}\right)  .
\end{equation}

We also introduce the auxiliary operators in Dirac space
\begin{subequations}
\begin{align}
D^{1/2(\pm)}_{B'BM} & \equiv  - \Gamma^{(\pm)} \left( \pm i\lambda +
  \frac{1 - \lambda}{m_{B'}^{} \pm m_B^{}}\, \slashed{\partial} \right)  ,  \\
D^{3/2(\pm)}_{\nu} & \equiv  \Gamma^{(\mp)} \partial_\nu \, , \\
D^{5/2(\pm)}_{\mu\nu} & \equiv - i \Gamma^{(\pm)} \partial_\mu \partial_\nu \, , \\
D^{7/2(\pm)}_{\mu\nu\rho} & \equiv  - \Gamma^{(\mp)} \partial_\mu \partial_\nu \partial_\rho \, ,
\label{eq:defps}
\end{align}
\end{subequations}
where $\Gamma^{(+)} \equiv \gamma_5^{}$ and $\Gamma^{(-)} \equiv 1$.
Here, $m_B^{}$ stands for the mass of the baryon $B$.
The parameter $\lambda$ has been introduced to interpolate between the 
pseudovector ($\lambda=0$) and the pseudoscalar ($\lambda=1$) couplings.
Note that in the above equation the order of the subscript indices in $D^{1/2(\pm)}_{B'BM}$ is important, i.e.,
$D^{1/2(\pm)}_{B'BM} \ne D^{1/2(\pm)}_{BB'M}$.

The effective Lagrangians for spin-1/2 hyperons $\Lambda$ and $\Sigma$ (or their resonances) are, then,
given by
\begin{subequations}\label{Lag_12}
\begin{align}
{\cal L}^{1/2 (\pm)}_{\Lambda NK} &=
g_{\Lambda NK}^{} \,\bar\Lambda  \left( D^{1/2(\pm)}_{\Lambda NK} \bar{K} \right) N
+ \mbox{H.c.} \, , \\
{\cal L}^{1/2(\pm)}_{\Sigma NK} &=
g_{\Sigma NK}^{} \,\bar{\bm{\Sigma}} \cdot \left( D^{1/2(\pm)}_{\Sigma NK}\bar{K}\right) \bm{\tau} N
+ \mbox{H.c.} \, , \\
{\cal L}^{1/2 (\pm)}_{\Xi\Lambda K_c} &=
g_{\Xi\Lambda K_c}^{} \,\bar\Xi  \left( D^{1/2(\pm)}_{\Xi\Lambda K}K_c \right) \Lambda
+ \mbox{H.c.} \, , \\
{\cal L}^{1/2 (\pm)}_{\Xi\Sigma K_c} &=
g_{\Xi\Sigma K_c}^{} \,\bar\Xi \,\bm{\tau} \left( D^{1/2(\pm)}_{\Xi\Sigma K}K_c \right) \cdot \bm{\Sigma}
+ \mbox{H.c.} \, ,
\end{align}
\end{subequations}
where the superscripts $\pm$ refer to the positive $(+)$ and negative $(-)$ relative parity of the baryons.
Flavor SU(3) symmetry relates the coupling constants among the members of the octet $J^P=1/2^+$
ground state baryons and $J^P=0^-$ pseudoscalar mesons and we have
\begin{subequations}\label{gB'BM}
\begin{align}
g_{\Lambda NK}^{} & = - g_8^{} \frac{1+2\alpha}{\sqrt{3}} \, , \\
g_{\Sigma NK}^{}  & = g_8^{} (1-2\alpha) \, , \\
g_{\Xi\Lambda K_c}^{} &  = - g_8^{} \frac{1-4\alpha}{\sqrt{3}} \, , \\
g_{\Xi\Sigma K_c}^{} &  = - g_8^{} \, ,
\end{align}
\end{subequations}
where the empirical values are $g_8^{} = g_{NN\pi}^{} = 13.26$ and $\alpha=0.365$, where $\alpha$
is the $F/D$ mixing parameter defined as $\alpha = F/(D+F)$.

For spin-3/2 hyperons, we have
\begin{subequations}\label{Lag_32}
\begin{align}
{\cal L}^{3/2 (\pm)}_{\Lambda NK} &=
\frac{g_{\Lambda NK}^{}}{m_K} \,\bar{\Lambda}^\nu  \left( D^{3/2(\pm)}_{\nu} \bar{K} \right) N  + \mbox{H.c.} \, , \\
{\cal L}^{3/2 (\pm)}_{\Sigma NK} &=
\frac{g_{\Sigma NK}^{}}{m_K} \,\bar{\bm{\Sigma}}^\nu \cdot \left( D^{3/2(\pm)}_{\nu} \bar{K} \right) 
\bm{\tau} N  + \mbox{H.c.} \, , \\
{\cal L}^{3/2 (\pm)}_{\Xi\Lambda K_c} &=
\frac{g_{\Xi\Lambda K_c}^{}}{m_K} \,\bar\Xi \left( D^{3/2(\pm)}_\nu K_c \right) \Lambda^\nu + 
\mbox{H.c.} \, , \\
{\cal L}^{3/2 (\pm)}_{\Xi\Sigma K_c} &=
\frac{g_{\Xi\Sigma K_c}^{}}{m_K} \,\bar\Xi \bm{\tau} \left( D^{3/2(\pm)}_\nu K_c \right) \cdot 
\bm{\Sigma}^\nu + \mbox{H.c.} \, ,
\end{align}
\end{subequations}
where $m_K^{}$ denotes the kaon mass.
For spin-5/2 hyperons \cite{Chang67a,MON11},
\begin{subequations}\label{Lag_52}
\begin{align}
{\cal L}^{5/2 (\pm)}_{\Lambda NK} &=
\frac{g_{\Lambda NK}^{}}{m^2_K} \,\bar{\Lambda}^{\mu\nu} \left( D^{5/2(\pm)}_{\mu\nu} \bar{K}
\right) N  + \mbox{H.c.} \, , \\
{\cal L}^{5/2 (\pm)}_{\Sigma NK} &=
\frac{g_{\Sigma NK}^{}}{m^2_K} \,\bar{\bm{\Sigma}}^{\mu\nu} \cdot \left( D^{5/2(\pm)}_{\mu\nu} 
\bar{K} \right) \bm{\tau} N + \mbox{H.c.} \, , \\
{\cal L}^{5/2 (\pm)}_{\Xi\Lambda K_c} &=
\frac{g_{\Xi\Lambda K_c}^{}}{m^2_K} \,\bar\Xi  \left( D^{5/2(\pm)}_{\mu\nu} K_c \right) \Lambda^{\mu\nu} + \mbox{H.c.} \, , \\
{\cal L}^{5/2 (\pm)}_{\Xi\Sigma K_c} &=
\frac{g_{\Xi\Sigma K_c}^{}}{m^2_K} \,\bar\Xi \bm{\tau} \left( D^{5/2(\pm)}_{\mu\nu} K_c \right) \cdot\bm{\Sigma}^{\mu\nu}
+ \mbox{H.c.} \, .
\end{align}
\end{subequations}
And, for spin-7/2 hyperons, we have~\cite{Chang67a,MON11}
\begin{subequations}\label{Lag_72}
\begin{align}
{\cal L}^{7/2 (\pm)}_{\Lambda NK} &=
\frac{g_{\Lambda NK}^{}}{m^3_K} \,\bar{\Lambda}^{\mu\nu\rho}  \left( 
D^{7/2(\pm)}_{\mu\nu\rho} \bar{K} \right) N  + \mbox{H.c.} \, , \\
{\cal L}^{7/2 (\pm)}_{\Sigma NK} &=
\frac{g_{\Sigma NK}^{}}{m^3_K} \,\bar{\bm{\Sigma}}^{\mu\nu\rho} \cdot \left( 
D^{7/2(\pm)}_{\mu\nu\rho} \bar{K} \right) \bm{\tau} N
+ \mbox{H.c.} \, , \\
{\cal L}^{7/2 (\pm)}_{\Xi\Lambda K_c} &=
\frac{g_{\Xi\Lambda K_c}^{}}{m^3_K} \,\bar\Xi  \left( D^{7/2(\pm)}_{\mu\nu\rho} K_c \right) \Lambda^{\mu\nu\rho} + \mbox{H.c.} \, , \\
{\cal L}^{7/2 (\pm)}_{\Xi\Sigma K_c} &=
\frac{g_{\Xi\Sigma K_c}^{}}{m^3_K} \,\bar\Xi \bm{\tau} \left( D^{7/2(\pm)}_{\mu\nu\rho} K_c
\right) \cdot\bm{\Sigma}^{\mu\nu\rho}
+ \mbox{H.c.} \, .
\end{align}
\end{subequations}
The coupling constants in the above Lagrangians corresponding to $\Lambda$ and $\Sigma$ resonances are
free parameters adjusted to reproduce the existing data.
For those resonances considered in the present work, they are given in Table~\ref{tab:para_g_contact}.

In the present work, all the meson-baryon-baryon vertices are obtained from the above Lagrangian.
In addition, each vertex is multiplied by an off-shell form factor given by
\begin{equation}
f(p^2_r, m_r, \Lambda_r) = \left(\frac{n \Lambda_r^4}{n \Lambda_r^4+(p^2_r-m^2_r)^2}\right)^n  ,
\label{formfactor}
\end{equation}
where $p^2_r$ and $m_r$ are the square of the 4-momentum and mass of the exchanged hyperon, respectively.
The cutoff parameter $\Lambda_r$ is chosen to have a common value $\Lambda_r \equiv \Lambda =900$~MeV
for all the $MBr$ vertices in order to keep the number of free parameters to a minimum.
Also, we choose $n=1$.

For the propagators of the dressed hyperons, we could in principle adopt the forms used in our previous
work~\cite{NH04,NH05,NOH08,MON11}.
However, in view of the limited amount of currently available data for the present reaction and the rather poor
quality of these data, here we adopt the simpler forms as given in the following.
For a spin-1/2 baryon propagator, we use
\begin{align}
S^{1/2}_r(p_r)
&= \frac{1}{\slashed{p}_r - m_r^{} + i\frac{\Gamma_r}{2}} \, ,
\label{eq:N12-prop}
\end{align}
where $\Gamma_r$ is the baryon width assumed to be constant, independent of energy.
For a stable (ground state) baryon, $\Gamma_r \to \epsilon$ with $\epsilon$ being positive
infinitesimal.

For spin-3/2, the dressed propagator reads in a schematic matrix notation
\begin{equation}
S^{3/2}_r(p_r)= \frac{1}{\slashed{p}_r - m_r^{} + i\frac{\Gamma_r}{2}}
\Delta \, ,
\label{N32-prop}
\end{equation}
where $\Delta$ is the Rarita-Schwinger tensor with elements
\begin{equation}
\Delta^{\mu\nu}=
-g^{\mu\nu}+\frac{1}{3}\gamma^\mu\gamma^\nu
  + \frac{2p^\mu p^\nu}{3m_r^2}
+\frac{\gamma^\mu
p^\nu-p^\mu\gamma^\nu}{3m_r^{}} \, .
\label{eq:RStensor}
\end{equation}
Similarly, the propagator for a spin-5/2 resonance is given by
\begin{equation}
S^{5/2}_r(p_r)= \frac{1}{\slashed{p}_r - m_r^{} + i\frac{\Gamma_r}{2}}
\Delta \, ,
\label{N52-prop}
\end{equation}
where the elements of $\Delta$ are~\cite{Chang67a}
\begin{eqnarray}
\Delta_{\alpha_1^{}\alpha_2^{}}^{\beta_1^{}\beta_2^{}}
&=& \frac12 \left( \bar{g}_{\alpha_1^{}}^{\beta_1^{}}
\bar{g}_{\alpha_2^{}}^{\beta_2^{}}
+ \bar{g}_{\alpha_1^{}}^{\beta_2^{}}
\bar{g}_{\alpha_2^{}}^{\beta_1^{}} \right)
-\frac15 \bar{g}_{\alpha_1^{}\alpha_2^{}} \bar{g}^{\beta_1^{}\beta_2^{}}
\nonumber \\ && \mbox{}
- \frac{1}{10} \left(
\bar{\gamma}_{\alpha_1^{}}^{} \bar{\gamma}^{\beta_1^{}}
\bar{g}_{\alpha_2^{}}^{\beta_2^{}}
+ \bar{\gamma}_{\alpha_1^{}}^{} \bar{\gamma}^{\beta_2^{}}
\bar{g}_{\alpha_2^{}}^{\beta_1^{}}
+ \bar{\gamma}_{\alpha_2^{}}^{} \bar{\gamma}^{\beta_1^{}}
\bar{g}_{\alpha_1^{}}^{\beta_2^{}} \right. \nonumber \\
&& \mbox{} \left. \qquad \quad \mbox{}
+ \bar{\gamma}_{\alpha_2^{}}^{} \bar{\gamma}^{\beta_2^{}}
\bar{g}_{\alpha_1^{}}^{\beta_1^{}} \right)~,
\label{eq:52tensor}
\end{eqnarray}
with
\begin{equation}
\bar{g}^{\mu\nu} \equiv g^{\mu\nu} - \frac{p^\mu p^\nu}{m_r^2} \, ,
\qquad
\bar{\gamma}^\mu \equiv \gamma^\mu - \frac{p^\mu \slashed{p}}{m_r^2}  \, .
\end{equation}
The propagator for a spin-7/2 resonance is given by
\begin{equation}
S^{7/2}_r(p_r)= \frac{1}{\slashed{p}_r - m_r^{} + i\frac{\Gamma_r}{2}}
\Delta \, ,
\label{N72-prop}
\end{equation}
where the elements of $\Delta$ are~\cite{Chang67a}
\begin{eqnarray}
\Delta_{\alpha_1^{}\alpha_2^{}\alpha_3^{}}
^{\beta_1^{}\beta_2^{}\beta_3^{}}
&=& \frac{1}{36} \sum_{P(\alpha),P(\beta)} \left(
\bar{g}_{\alpha_1^{}}^{\beta_1^{}} \bar{g}_{\alpha_2^{}}^{\beta_2^{}}
\bar{g}_{\alpha_3^{}}^{\beta_3^{}}
-\frac37
\bar{g}_{\alpha_1^{}}^{\beta_1^{}} \bar{g}_{\alpha_2^{}\alpha_3^{}}^{}
\bar{g}^{\beta_2^{}\beta_3^{}} \right. \nonumber \\
&& \mbox{} -  \left. \frac37
\bar{\gamma}_{\alpha_1^{}}^{} \bar{\gamma}^{\beta_1^{}}
\bar{g}_{\alpha_2^{}}^{\beta_2^{}} \bar{g}_{\alpha_3^{}}^{\beta_3^{}}
+ \frac{3}{35}
\bar{\gamma}_{\alpha_1^{}}^{} \bar{\gamma}^{\beta_1^{}}
\bar{g}_{\alpha_2^{}\alpha_3^{}}^{} \bar{g}^{\beta_2^{}\beta_3^{}} \right) ,
\nonumber\\
\label{eq:72tensor}
\end{eqnarray}
and the summation runs over all possible permutations of
$\{\alpha_1^{}, \alpha_2^{}, \alpha_3^{} \}$
and of $\{\beta_1^{}, \beta_2^{}, \beta_3^{} \}$.

To avoid an ambiguity in the relative phase between $M_s+M_u$ and $M_c$ in
Eq.~(\ref{eq:BS-approx}), we provide here the explicit expressions for the
amplitudes $M_s$ and $M_u$ for the $\Lambda(1116)$ exchange in the $K^-(q)
+N(p)\rightarrow K^+(q')+\Xi^-(p')$
 reaction, i.e.,
\begin{subequations}
\begin{align}
M_{s}^{\Lambda} &= \bar{u}_\Xi^{} (\bm{p'}) \, \Gamma^s_{\Xi^-K^+\Lambda }(q')
\, S^{1/2}_{\Lambda}(p_s) \,
 \Gamma^s_{\Lambda  K^- N}(q) \,u_N^{}(\bm{p})~,
   \\
M_{u}^{\Lambda} &= \bar{u}_\Xi^{} (\bm{p'}) \, \Gamma^u_{\Xi^- K^-\Lambda}(q)
\, S^{1/2}_{\Lambda}(p_u) \,
\Gamma^u_{\Lambda K^+ N}(q')\, u_N^{} (\bm{p}) ~,
 \end{align}
 \end{subequations}
where the nucleon index $N$ stands for the proton and $\Lambda$ stands for
$\Lambda(1116)$; the baryon Dirac spinors are normalized covariantly,
$\bar{u}_B^{} u_B^{} = 1$, the intermediate four-momenta are $p_s=p+q$ and $p_u=p-q'$,
and the vertices are given as
\begin{subequations}
\begin{align}
\Gamma^s_{\Lambda  K^- N}(q)&=g_{\Lambda N K}^{}\ \gamma^5\left(\lambda-
\frac{1-\lambda}{m_\Lambda^{} + m_N^{}}\slashed{q}\right)   f_s~, \\
\Gamma^s_{\Xi K^+\Lambda}(q')&=g_{\Xi\Lambda K}^{}\ \gamma^5\left(\lambda+
\frac{1-\lambda}{m_\Xi^{}+m_\Lambda^{}}\slashed{q}'\right)   f_s~, \\
\Gamma^u_{\Lambda K^+N}(q')&=g_{\Lambda N K}^{}\ \gamma^5\left(\lambda
+\frac{1-\lambda}{m_\Lambda^{}+m_N^{}}\slashed{q}'\right)   f_u~, \\
\Gamma^u_{\Xi K^- \Lambda }(q)&=g_{\Xi\Lambda K}^{}\ \gamma^5\left(\lambda
-\frac{1-\lambda}{m_\Xi^{} + m_\Lambda^{}}\slashed{q}\right)   f_u~,
\end{align}
\end{subequations}
where $\lambda$ describes the linear interpolation between pseudoscalar
($\lambda=1$) and pseudovector ($\lambda=0$) couplings; the off-shell form
factors are given by [see Eq.~(\ref{formfactor})]
 \begin{subequations}
\begin{align}
  f_s &= f(p_s^2,m_\Lambda,\Lambda_\Lambda) \, ,  \\
  f_u &= f(p_u^2,m_\Lambda,\Lambda_\Lambda) \, ,
\end{align}
\end{subequations}
and the values of $g_{\Lambda N K}^{}$, $g_{\Xi\Lambda K}^{}$, and $\lambda$ are
found in Table~\ref{tab:para_g_contact}.

%%%%%%%%%%%%%%%%%%%%%%%%%%%%%%%%%%%%%%%%%%%%%%%%

\end{document}